\documentclass[aps,showkeys,longbibliography]{revtex4-2}
\usepackage{graphicx}
\usepackage{subfigure}
\usepackage{color}
\usepackage{amsmath,amssymb,mathtools}
\usepackage{xcolor}
\usepackage[colorlinks=true,allcolors=blue]{hyperref}
\usepackage{bbold}
\usepackage[T1]{fontenc}
\usepackage{etoolbox}
\usepackage{physics}
\usepackage{braket}
\usepackage{bm,bbm}
\usepackage{braket}
\usepackage{caption}
\usepackage{float}

\hypersetup{colorlinks=true,linkcolor=blue,citecolor=blue,urlcolor=blue}

\makeatletter
\newcommand{\pushright}[1]{\ifmeasuring@#1\else\omit\hfill$\displaystyle#1$\fi\ignorespaces}
\newcommand{\pushleft}[1]{\ifmeasuring@#1\else\omit$\displaystyle#1$\hfill\fi\ignorespaces}
\makeatother

\def\vk {\bm{k}}
\def\vd {\bm{d}}

\def\vs {\bm{\sigma}}
\def\vt {\bm{\tau}}
\def\vdt {\bm{\delta}}

\def\s {s}

\def \dd {\mathrm{d}}

\def\ic{\mathrm{i}}

\def\bc{\begin{center}}
\def\ec{\end{center}}
\def\bi{\begin{itemize}}
\def\ei{\end{itemize}}
\def\ba{\begin{array}}
\def\ea{\end{array}}

\DeclareMathOperator{\sgn}{sgn}

\newcommand{\be}{\begin{equation}}
\newcommand{\ee}{\end{equation}}

\newcommand{\ra}{\rangle}

\begin{document}

\title{Real and momentum space analysis of topological phases in 2D d-wave altermagnets}

\author{Manuel Calixto}
\email{calixto@ugr.es}
\affiliation{Department of Applied Mathematics and Institute Carlos I of Theoretical and Computational Physics, University of  Granada, Granada 18071, Spain}




\begin{abstract}
\noindent \textbf{Abstract:} 
Altermagnetism has recently emerged as a third fundamental branch of magnetism, combining the vanishing net magnetization of antiferromagnets with the high-momentum-dependent spin splitting of ferromagnets. This study provides a comprehensive real- and momentum-space analysis of topological phases in two-dimensional d-wave altermagnets. By employing a tight-binding Hamiltonian, we characterize the topological phase transition occurring at a critical intra-sublattice hopping strength ($t_a^C$). We examine the emergence of Dirac nodal points and the resulting Berry curvature singularities, supported by a visual analysis of pseudospin texture winding. Crucially, we analize spin splitting, effective altermagnetic strength, and investigate the transport implications of these phases, uncovering giant conductivity anisotropy and spin-dependent ``steering''  effects driven by group velocity distribution across the Fermi surface. Beyond bulk properties, we analyze the edge state topology in ribbon geometries through the lens of information-theoretic markers like fidelity-susceptibility and  inverse participation ratio, offering a robust alternative to traditional Chern number calculations, without relying on translational symmetry. Our results demonstrate that the hybridization of edge states in ultra-narrow nanoribbons opens a controllable energy gap, a feature we exploit to propose a novel topological altermagnetic field-effect transistor design where ballistic and spatially spin-polarized transport can be electrostatically gated. This work establishes a theoretical and information-theoretic framework for ``edgetronics'' in altermagnetic materials, paving the way for next-generation, high-speed spintronic and ``spin-splitter'' logic devices and architectures.

\end{abstract}

\keywords{Altermagnetism, Topological Phase Transition, d-wave symmetry, Spin-splitting, Edgetronics, Nanoribbon FET, Quantum Information}

\maketitle 

\section{Introduction}

For nearly a century, our understanding of magnetic order was built upon a fundamental binary: ferromagnetism (FM), characterized by parallel spin alignment and a net macroscopic magnetization, and antiferromagnetism (AFM), where antiparallel spins cancel each other out, resulting in a zero net magnetic moment. As noted in a recent editorial by Igor Mazin \cite{Mazin2022_Editorial}, altermagnetism (AM)  has emerged as a ``new punch line'' (a third branch) in fundamental magnetism, disrupting this traditional dichotomy by introducing a third magnetic phase that combines the most desirable properties of both worlds.

The formalization of AM as a distinct phase was established in the seminal work by \v{S}mejkal, Sinova, and Jungwirth \cite{Smejkal2022_Theory}. This theoretical framework has mapped out an extensive ``research landscape'' \cite{Smejkal2022_Landscape}, identifying numerous candidate materials across metals, insulators, and semiconductors. They demonstrated that certain collinear magnets possess a unique symmetry where opposite magnetic sublattices are connected by crystal rotations rather than the translations or inversions typical of conventional AFMs. This ``nonrelativistic spin and crystal rotation symmetry'' results in a vanishing net magnetization (like an AFM) but allows for large, momentum-dependent spin splitting in the electronic bands (like a FM), even without considering spin-orbit coupling.

Indeed, the most striking signature of an AM is the  Kramers spin degeneracy lifting in momentum space, experimentally confirmed in  \cite{Krempasky2024_MnTe}. Traditional AFMs maintain degenerate spin-up and spin-down bands due to the combination of time-reversal and inversion symmetry, whereas AMs break this degeneracy. In colloquial terms, spin polarization is the ``secret sauce''. The significance of AM lies in its potential for spintronics. The spin polarization in AMs is not a weak effect because it is protected by the crystal lattice symmetry. Spin polarization allows the material to behave like a FM in transport experiments while acting like an AFM in terms of magnetic stability. In other words,  AMs combine the best of both worlds. Usually, spin polarization is a trade-off. To get spin-polarized currents, you use Ferromagnets (FM), but they produce stray magnetic fields that interfere with neighboring devices and are limited by slow dynamics. We use AFMs to avoid stray fields and low dynamics, but they usually have degenerate (non-polarized) bands, making them harder to read, manipulate, and use for electronic signals. AMs break  this trade-off. Since the spin polarization in an AM is directionally dependent (e.g., d-wave, g-wave or p-wave symmetries \cite{10.21468/SciPostPhys.18.3.109} describing the shape of the spin splitting in momentum space), an AM acts as a ``spin filter'' that depends on the electron's velocity. We will show later in Section \ref{condsec}  how, in a $\dd_{x^2-y^2}$-wave AM, electrons traveling along the $x$-axis could have mostly spin up, while those along the $y$-axis could have mostly spin down.  In short, AMs offer the high-speed, robust nature of AFMs while providing the spin-polarized currents necessary for device functionality.

Recently, an exhaustive classification of bulk minimal models for altermagnetism across various space groups was established by Roig et al. \cite{Roig_PhysRevB.110.144412}. Building upon the foundational 2D $\dd$-wave Hamiltonians identified in such classifications, our study explicitly breaks translational symmetry to investigate the resulting topological phase transitions and the real-space emergence of edge states in nanoribbon geometries, a regime where bulk classifications must be supplemented by finite-size topological markers (edge state localization, hibridization gap, fidelity-susceptibility, etc.).

In terms of experimental observation, two prototypical materials have provided the ``smoking gun'' evidence for this exotic topological phase of nature. First, research on  {MnTe}  using angle-resolved photoemission spectroscopy, provided direct experimental confirmation of AM band splitting  \cite{Krempasky2024_MnTe,Mazin2023_Notes}. Second, in the metallic regime, ruthenium dioxide (RuO$_{2}$) has served as a critical test bench. Early experimental evidence for AM in RuO$_{2}$ came from the observation of a ``Crystal Hall Effect''  \cite{Feng2022_RuO2_AHE,Fedchenko2024_RuO2_ARPES}. While materials like RuO$_2$, MnTe, and MnF$_2$ are canonically proposed as $\dd$-wave and g-wave altermagnets, we must note that their exact magnetic ground states and the presence of altermagnetic splitting are currently subjects of active experimental and theoretical debate \cite{PhysRevLett.133.176401,PhysRevLett.134.226702,zeng2025nonaltermagneticspintexturemnte}. Our minimal tight-binding model is designed to capture the characteristic topological features of symmorphic tetragonal $\dd$-wave altermagnetic phases (such as those compatible with Space Group 123), providing a clean theoretical framework to study finite-size effects with information-theoretic tools that could be extended to more general cases. 

Cold atoms trapped in optical lattices represent an alternative route for AM realization. They represent the ideal testbed for understanding the interplay between electronic correlations and the emergence of AM in many fermion systems, a two-dimensional version of the Hubbard model suitably modified for hosting altermagnetic states that can be simulated via nowadays available cold-atom technology \cite{Das_PhysRevLett.132.263402}. 
The list of references is enormous, with considerable growth in recent years.

Prospects in giant magnetoresistance is also promising because  AMs allow the creation of magnetic tunnel junctions due to massive changes in resistance depending on the relative orientation of two AM layers. AMs may also improve the efficiency of Spin-Orbit Torque Magnetic Random-Access Memories (SOT-MRAM).
While conventional SOT-MRAM relies on the spin-Hall effect in heavy metals \cite{PhysRevB.78.212405,Miron2011,doi:10.1126/science.1218197}, AMs offer a newly discovered mechanism: the Spin-Splitter Torque (SST) \cite{PhysRevLett.126.127701}. Recent experiments in RuO$_2$ have demonstrated that this intrinsic torque is highly efficient \cite{Bai2021ObservationOS}, paving the way for single-layer, low-power, ultra-fast memory devices that combine the readout capabilities of FMs with the Terahertz dynamics of AFM \cite{Nguyen2024}.

The present study takes a theoretical approach while keeping an eye on possible physical applications. Our goal is to delve deeper into the internal structure of the spectrum and wave functions to better understand the nature of the topological phase transition (TPT) in AMs. The article also aims to be instructive, providing sufficient detail to be as self-contained as possible, so that it can be read by an audience that is not necessarily an expert in the subject. The reason for doing so is that, we also follow an information-theory perspective, which may not be the usual approach for readers coming from condensed matter physics, and vice versa. The article is organized as follows. In Section \ref{latticesec} we introduce the model, discuss the real-space crystal structure and the momentum-space band structure, as well as the corresponding tight-binding Hamiltonians and their diagonalization. Section \ref{berrysec} is dedicated to the discussion of nodal/Dirac points and a Berry phase analysis to visualize and highlight the TPT that occurs in AMs for a critical value of the intra-sublattice AM hopping strength $t_a$.  In Section \ref{condsec} we study the spin spliting and  effective altermagnetic strength, together with the group velocity distribution across the Fermi surface, and conductivity anisotropy profiles. The information-theoretic concept of fidelity-susceptibility is introduced in Section \ref{realspacesec} to characterize TPTs as an alternative to traditional topological quantum (Chern, Pontryagin, winding, etc) numbers. This Section is also devoted to  a real-space crystal structure analysis, together with a calculation and interpretation  of the spatial segregation texture (or chirality) and  global and local density of states in line with the spin-splitting and conductivity anisotropy pictures. In Section \ref{gapsec} we study the appearance of a hybridization gap in finite nanoribbons; this robust and controllable gap can be used to design a novel altermagnetic topological field-effect transistor (FET), where ballistic spin transport can be electrostatically gated. Another information-theoretic concept, called inverse participation ratio (IPR) and linked to the R\'enyi entropy of order two, is introduced in Section \ref{ribbonsec} to analyze the edge state structure and localization in an infinite strip geometry. Finally, Section \ref{conclusec} is left for conclusions and future directions.

\section{$\dd$-wave Altermagnet Lattice Structure and Hamiltonian Model}\label{latticesec}

\begin{figure}[h]
	\begin{center}
		\includegraphics[width=0.55\columnwidth]{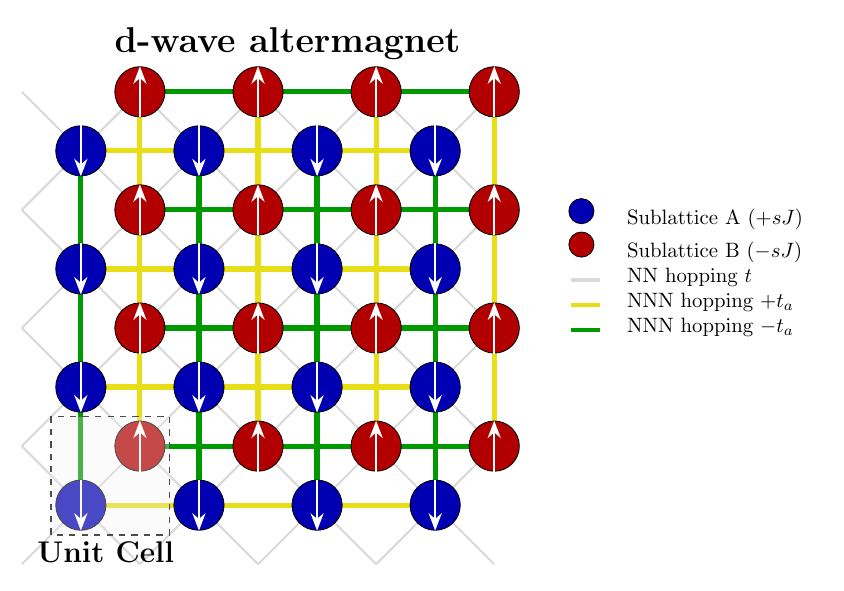}
			\includegraphics[width=0.35\columnwidth]{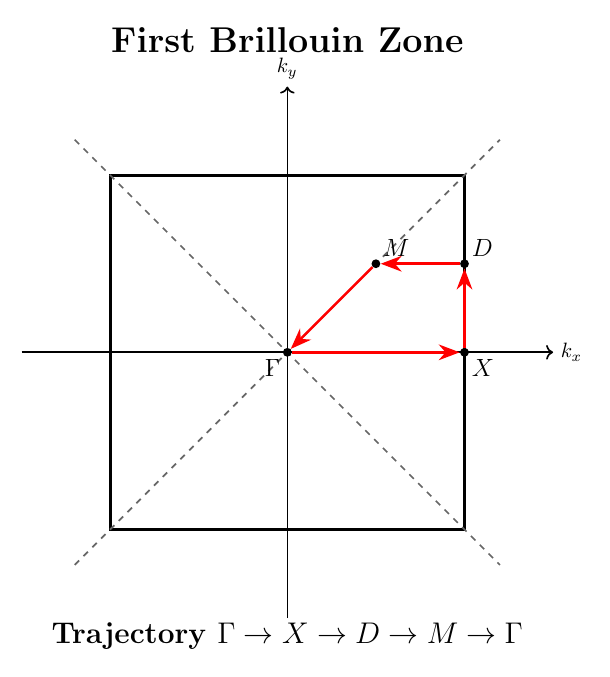}
	\end{center}
	\captionsetup{justification=raggedright, singlelinecheck=false}
	\caption{Real-space representation of a 2D $\dd$-wave AM  lattice, its first Brillouin zone and a particular closed path.}
	\label{fig_AMlattice}
\end{figure}

We start defining a minimal tight-binding Hamiltonian that captures the essential symmetry of $\dd$-wave altermagnetism. This model serves as a controllable theoretical laboratory (extensively studied in the context of both transition metal oxides and ultracold atom realizations) to investigate the interplay between exchange splitting and anisotropic hopping, establishing the foundational parameters used throughout the subsequent topological and transport analyses.

In Figure \ref{fig_AMlattice} we represent the real-space crystal structure of  a 2D lattice $\dd$-wave altermagnet. The sublattice $A$  (blue) and sublattice $B$ (red) possess opposite magnetic moments $-m$ and $m$ represented by down $\downarrow$ and up $\uparrow$ white arrows. The sublattices $A$  and $B$  then act as an antiferromagnet, exerting opposite exchange fields on the spin $\s=\pm 1$ itinerant electrons.
Due to the strong on-site exchange interaction (Hund's coupling), an itinerant electron spin aligns with the local magnetic moment to minimize its exchange energy. In the Hamiltonian, this results in a staggered exchange potential energy of $+\s J$ and $-\s J$ for $A$ and $B$ type sublattices, respectively, where $J$ represents the effective spin-splitting energy (exchange field strength). The gray lines represent nearest-neighbor (NN) hopping $t$ between $A$ and $B$ sites, while the  yellow and green lines represent the characteristic altermagnetic next-nearest-neighbor (NNN) hopping amplitudes $+t_{a}$ and $-t_{a}$, respectively. Let us denote by $\vk=(k_x,k_y)$ the wave vector and by $\vt=(\tau_x,\tau_y,\tau_z)$ the Pauli matrices acting on the lattice (pseudo-spin) sector $(A,B)$. The Bloch  Hamiltonian for spin $\s$ electrons in momentum space adopts the form $H_\s(\vk)=\vd_\s(\vk)\cdot\vt$, where the three-vector $\vd_\s=(d_x^\s,d_y^\s,d_z^\s)$ plays the role of an ``effective Zeeman field in pseudospin space''.

The NN interaction contributes to the Bloch Hamiltonian $H_\s(\vk)$ through a $d_x^\s\tau_x+d_y^\s\tau_y$ term with
\be
d_x^\s(\vk)-id_y^\s(\vk)=t\sum_{\vdt}e^{i\vk\cdot\vdt},
\ee
where $\vdt=a(\pm 1/2, \pm 1/2)$ are the four NN vector hoppings, so that  
\be
d_y^\s=0,\; d_x^\s(\vk)=4 t \cos \left(\frac{ak_x}{2}\right) \cos \left(\frac{ak_y}{2}\right).
\ee
Note that, since $d_y^\s=0$, the Bloch Hamiltonian $H_\s(\vk)$ is real and PT symmetric. This can be unconvenient for later Berry curvature analysis or for numerical/computational calculations in real space. Therefore, we also use the unitarily/gauge equivalent configuration
\be
\vdt=a(0,0), a(-1,0), a(0,-1), a(-1,-1)\rightarrow d_x^\s(\vk)-id_y^\s(\vk)=4 t \cos \left(\frac{ak_x}{2}\right) \cos \left(\frac{ak_y}{2}\right)e^{-i\phi(\vk)},\label{gauge}
\ee
where $\phi(\vk)=a(k_x+k_y)/2$ is a phase that does not change the Hamiltonian spectrum.

We shall measure energy in hopping $t$ units (eV scale) and length in lattice constant $a$ units (nm scale), to maintain generality. In practical calculations this means to set $t=1, a=1$. In particular, from now on we will write $a\vk\to\vk$ for simplicity.

The NNN AM interaction and the exchange potential energy $\s J$ both contribute to the $d_z^\s\tau_z$ Hamiltonian part as follows
\be
d_z^\s(\vk)=\s J+\sum_{\vdt'}t_{\vdt'} e^{i\vk\cdot\vdt'},
\ee
where $\vdt'$ denote the NNN vector hoppings and $t_{\vdt'}$ the corresponding NNN hopping amplitudes. The local environment permits generic hoppings $t_x$ and $t_y$ for sublattice A. Recently, a comprehensive symmetry analysis by Roig et al. \cite{Roig_PhysRevB.110.144412} demonstrated that for $\dd$-wave altermagnets (such as those in tetragonal space groups), the specific Wyckoff positions occupied by magnetic atoms are connected by a $C_4$ rotation. This symmetry rigidly dictates that the local hopping environment of sublattice A, when rotated by $90^\circ$, must perfectly map onto sublattice B. Due to the $C_4$ rotation symmetry connecting the sublattices, sublattice B must have hoppings $t_y$ (along $x$) and $t_x$ (along $y$). The dispersion for A is then $E_A \sim t_x \cos k_x + t_y \cos k_y$, and the dispersion for B is $E_B \sim t_y \cos k_x + t_x \cos k_y$. The resulting kinetic energy can be algebraically decomposed into: 
$$E_{A,B} \sim \frac{t_x + t_y}{2}(\cos k_x + \cos k_y) \pm \frac{t_x - t_y}{2}(\cos k_x - \cos k_y).$$
The first (isotropic) term is just a normal, spin-degenerate band. The second (antisymmetric) term is the $\dd$-wave altermagnetic splitting. We shall chose $t_x = -t_y = t_a$ merely to set the isotropic, spin-degenerate background to zero, allowing us to focus entirely on the $\dd$-wave splitting term, that is, to isolate the purely altermagnetic physics. This symmetry imposes that, for a given sublattice,  the hopping $t_a$ along one axis must have the opposite sign to the hopping along the orthogonal axis, effective mimicking the $\dd$-orbital symmetry (or electronic configurations) for the azimuthal quantum number $l=2$ in the atom (see later in Figure \ref{fig_SpinConductivityTensor} for conductivity polar plots).  Therefore, the topology only depends on $(t_x - t_y) \neq 0$; actually,  as long as $t_x \neq t_y$, the $\dd$-wave form factor $\sim(\cos k_x - \cos k_y)$ perfectly survives, and the topological conclusions, including the existence of the Dirac dispersions, remain entirely robust.  Summarizing,  the $d_z^\s\tau_z$ Hamiltonian part reduces to
\be
d_z^\s(\vk)=\s J+t_a (e^{-i k_x}+e^{i k_x}-e^{-i k_y}-e^{i k_y})=\s J+2 t_a (\cos (k_x)-\cos (k_y)).\label{dzAM}
\ee
Thus, our tight-binding formulation represents the exact 2D projection of the bulk non-relativistic symmetry classes identified in recent literature, providing a rigorous foundation to explore its finite-size and topological consequences. Moreover, our model is also  essentially the standard minimal tight-binding model for $\dd$-wave AM, which has been recently established in the literature and proposed for realizations in ultracold atomic optical lattices \cite{Das_PhysRevLett.132.263402} using a Hubbard model with anisotropic next-nearest neighbor hopping (we are discarding on-site Hubbard interactions). Building upon this foundational bulk model, our focus will shift to its behavior in finite nanoribbon geometries later in Section \ref{realspacesec}.

The AM 2D lattice structure depicted in \ref{fig_AMlattice} can be rotated an angle $\beta$ (the angle between the $x$-axis and the nodal lines), so that the low momentum (long wavelength) expansion of the  $d_z$ Hamiltonian component gives a magnetic order \cite{Fukaya_2025}
 $$2 k_xk_y \sin(2\beta)+(k_x^2 - k_y^2)\cos(2\beta).$$
The cases $\beta=0$ and $\beta=\pi/4$ are referred in the literature to as $\dd_{x^2-y^2}$- and $\dd_{xy}$-wave AM, respectively, 
 imitating two of the five  $\dd$ orbitals $\{\dd_{x^2-y^2}, \dd_{xy}, \dd_{xz}, \dd_{zx}, \dd_{z^2}\}$ (or electronic configurations) for the azimuthal quantum number $l=2$ in the atom. The $\dd_{xy}$ expression serves as a formal mathematical transformation under a $45^\circ$ rotation of the reference frame, rather than representing an independent, stable 2D crystal symmetry class. While the bulk is identical for $\dd_{xy}$ and $\dd_{x^2-y^2}$ configurations, the edge physics is drastically different depending on whether the ribbon is cut along [100] ($d_{x^2-y^2}$ aligned) or [110] ($d_{xy}$ aligned). This fact will be relevant when discussing the ``FET proposal for nanoribbons'' in Section \ref{gapsec}, where we show that the [110] cut allows for a desired spin-splitter effect due to the group velocity anisotropy.
 
 While our current model maintains $d_y^s = 0$ (preserving specific spin symmetries), the introduction of Rashba or Dresselhaus spin-orbit coupling (SOC) in experimental setups would break these symmetries, potentially canting the spin texture and mixing the spin-polarized edge channels  \cite{Roig_PhysRevB.110.144412}. The interesting consequences of SOC are left for future studies.

The Bloch Hamiltonian $H_\s=\vd_\s\cdot\vt$ eigenvalues [conduction (+) and valence (-) energy bands] and eigenstates (two-spinors) are simply as follows
\be
E^\pm_\s=\pm |\vd_\s|,\quad
|\psi^\s_\pm\rangle=\left(\ba{c} \psi_A^\s\\ \psi_B^\s\ea\right)_\pm= \left(\ba{c}\frac{d_z^\s\pm|\vd_\s|}{d_x^\s+id_y^\s}\\ 1\ea\right).\label{eigenH}
\ee
The valence band minimum and the conduction band maximum energies are attained at the $\Gamma$ point $\vk=(0,0)$ with values
\be
E_\s^\pm(\Gamma)=\pm\sqrt{J^2 \s^2+16 t^2}.\label{cbme}
\ee
We shall fix $J=t$ as an illustrative case (rather than a material-specific one) for numerical calculations along the paper, which gives $E_\s^\pm(0,0)\simeq 4.12$ in $t$ units. $J=t$ is chosen to maximize the visibility of the band-splitting effects in the figures. Reference \cite{Smejkal2022_Theory} (supplementary material) fits the DFT bands of KRu$_4$O$_8$ to $t = 0.1, J = 0.2$ and $t_a=0.075$. Therefore, while $J=t$ means  ``strong coupling'', it is the correct order of magnitude to reflect the strong exchange splitting observed in candidate altermagnetic materials like RuO$_2$, where the spin splitting reaches magnitudes of $\sim 1$~eV. Our conclusions on the existence of the topological phase, the associated edge states, etc., are robust for our illustrative choice $J=t$.

In the next sections, we analyze the structure and evolution of the valence band maximum (VBM) and conduction band minimum (CBM) Hamiltonian eigenstates as a function of the AM hopping $t_a$. For $t_a\leq J/4$ (low AM hopping strength compared to exchange field strength), the energies attained at $\vk=(0,\pm\pi)$ (in inverse lattice constant $a^{-1}$ units) for $\s=-1$ and at $\vk=(\pm\pi,0)$ for $\s=1$ give a band gap of
\be
E_g = E_{\text{CBM}} - E_{\text{VBM}}=2(J-4t_a),\quad t_a\leq J/4.
\ee
The topological AM structure around the critical point, 
\be
t_a^C=J/4,\label{tacrit}
\ee
where the gap closes, $E_g=0$, is analyzed in the next section.

\section{Nodal/Dirac Points and Berry Phase Analysis}\label{berrysec}

In this section, we review the standard momentum-space topological properties of the bulk $\dd$-wave altermagnetic phase. We characterize the gapless Dirac nodal semimetal regime and evaluate its local topological invariants, specifically the quantized nodal Berry phases and winding numbers. While these conventional bulk metrics are well-established for infinite, translationally invariant systems, they can become ill-defined or inaccessible in realistic, strictly confined nanoribbon geometries where translation symmetry is broken. Therefore, the analysis presented here serves primarily as a foundational comparative baseline. Establishing these bulk nodal properties allows us to directly benchmark the core conceptual advance of our work: the application of real-space, device-oriented metrics—such as fidelity susceptibility and the IPR developed in Sections \ref{realspacesec} and \ref{ribbonsec}, which provide a robust alternative for diagnosing quantum phase transitions in finite-size topological systems.

To identify the transition between trivial and topological regimes, we locate the Dirac points where the energy gap closes. Analyzing these nodal points and the associated Berry phase allows us to visualize the pseudospin texture winding, providing a clear momentum-space signature of the topological phase transition.

\begin{figure}[ht]
\begin{minipage}[b]{0.45\columnwidth}
        \includegraphics[width=\textwidth]{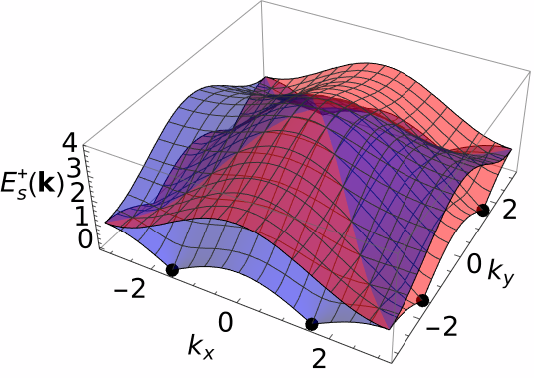}
        \centerline{(a)} 
    \end{minipage}
    \hfill
    \begin{minipage}[b]{0.45\columnwidth}
        \includegraphics[width=\textwidth]{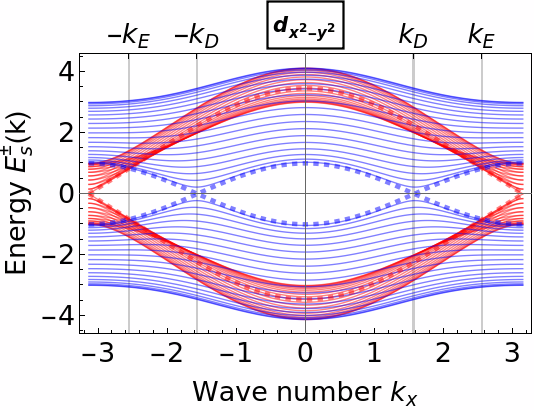}
        \centerline{(b)} 
    \end{minipage}
	\captionsetup{justification=raggedright, singlelinecheck=false}
	\caption{(a) Energy dispersion relation $E^+_s(k_x,k_y)$ as a function of $k_x, k_y\in [-\pi,\pi]$, for conduction ($+$) states for  spin $\uparrow$ (red) and spin $\downarrow$ (blue) electrons in the TI phase ($t_a>t_a^C$). Dirac points are marked with black dots. (b) Cross sections  of $E^\pm_s(k_x,k_y)$ for  $k_y=-n\pi/20, n=0,1,\dots,20$. Edge states  closing the gap at the Dirac points $k_D$ are marked by thick dashed curves: $E^\pm_\downarrow(k_x,-\pi)$ (blue) and $E^\pm_\uparrow(k_x,-k_D)$ (red). Model parameters $J=1, t_a=0.5$ (NN hopping $t$ and lattice constant $a$ units).}
	\label{fig_BandStructure}
\end{figure}

The analytical conditions for the emergence of Dirac points (in the bulk spectrum of two-dimensional Hubbard model adapted to host altermagnetic states) were expertly derived by Del Re \cite{DelRe7PRR7} in the context of Hartree-Fock and  dynamical mean-field theory studies. Here, we briefly re-derive these locations to explicitly connect them to the Berry curvature singularities, which we will later use to benchmark our real-space topological markers.

The energy gap $E_g=2|\vd_s|$ closes only when $d_x^\s=0=d_z^\s$.  The off-diagonal hopping term is zero when
\be
d_x^\s(\vk)=4t \cos(k_x/2) \cos(k_y/2) = 0\Rightarrow k_x=\pm\pi \quad \mathrm{or}\quad k_y=\pm\pi,
\ee
which defines the nodal lines. For $k_y = \pm\pi$,
\be
d_z^\s(\vk)=\s J + 2t_a(\cos(k_x) - \cos(\pi)) = 0\Rightarrow k_x = \pm \arccos\left(\frac{J}{2t_{a}} - 1 \right)\equiv \pm k_D,\ee
which defines the Dirac points $\vk_{D}^\pm=(\pm k_D,-\pi)$ [and equivalent $(\pm k_D,\pi)$] for $t_a>t_a^C$  and $\s=-1$, with $t_a^C$ the critical point already defined in \eqref{tacrit}. We assume that $J, t_a\geq 0$. For $k_x=\pm\pi$ and $\s=1$, the Dirac points are $\vk_{D}'^\pm=(-\pi,\pm k_D)$ [and the equivalent $(\pi,\pm k_D)$].

In Figure \ref{fig_BandStructure} we represent the Hamiltonian spectrum $E^\pm_\s(\vk)$ as a function of $(k_x,k_y)$ for the spin-up (red) and spin-down (blue) electrons in the TI phase ($t_a>t_a^C$). The Hamiltonian spectrum is equivalent under the combined exchange of $k_x\leftrightarrow k_y$ and $\uparrow\leftrightarrow \downarrow$ (i.e., blue$\leftrightarrow$red). Edge states develop above the critical point $t_a^C$, closing the gap at the Dirac points $\pm k_D$. The thick dashed curves correspond to $E^\pm_\downarrow(k_x,-\pi)$ (blue) and $E^\pm_\uparrow(k_x,-k_D)$ (red). A crossing of both curves occurs at
\be
k_E= \arcsec\left(\frac{2 \left(t^2+2 t_a^2\right)}{J t_a -2 t^2-4 t_a^2}\right),
\ee
($E$ makes reference to ``edge'') where spin-up and spin-down edge Hamiltonian eigenstates degenerate. Dirac ($k_D$) and edge crossing ($k_E$) points are marked by vertical grid lines in Figure \ref{fig_BandStructure}(b).

Defining the Fermi velocities
\be
v_1=\sqrt{(4t_a-J)J}/\hbar,\quad  v_2=t\sqrt{J/t_a}/\hbar,\quad t_a>t_a^C,
\ee
the first-order energy dispersion expansions around the Dirac points $\vk_{D}^\pm$ and $\vk_{D}'^\pm$ are
\be
E_{-1}^\pm(q_x,q_y)=\pm\hbar\sqrt{v_1^2q_x^2+v_2^2 q_y^2},\quad \mathrm{and}\quad
E_{1}^\pm(q'_x,q'_y)=\pm\hbar\sqrt{v_2^2q_x'^2+v_1^2 q_y'^2},
\ee
with $\bm{q}=\vk-\vk_{D}^\pm$ and $\bm{q}'=\vk-\vk_{D}'^\pm$. The anisotropy in the Fermi velocity leads to directional dependence in the transport, which  will be discussed in Section \ref{condsec}.

\begin{figure}[h]
    \begin{minipage}[b]{0.48\columnwidth}
        \includegraphics[width=\textwidth]{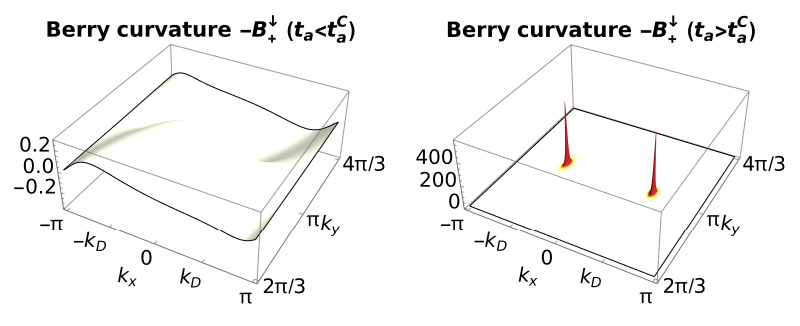}
        \centerline{(a)} 
    \end{minipage}
    \hfill
    \begin{minipage}[b]{0.48\columnwidth}
        \includegraphics[width=\textwidth]{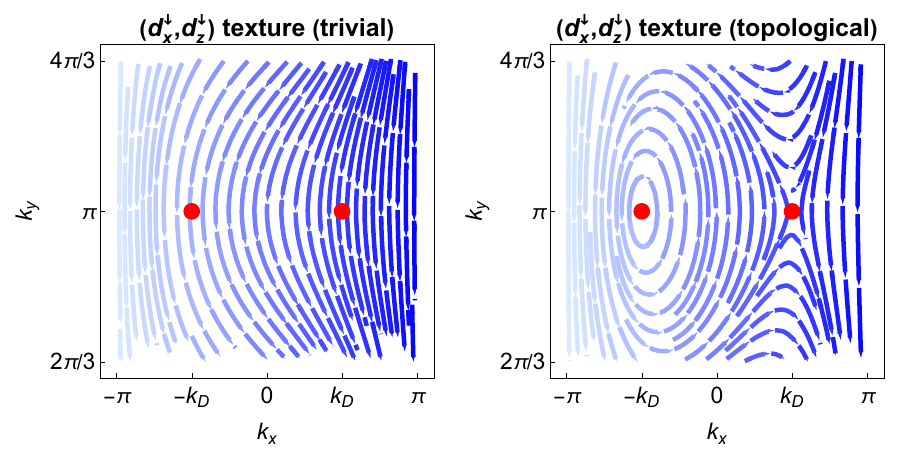}
        \centerline{(b)} 
    \end{minipage}
    \captionsetup{justification=raggedright, singlelinecheck=false}
    \caption{Berry curvature (a) and pseudo-spin texture (b) as a function of $\vk=(k_x,k_y)$ for spin $\s=-1$ in the trivial phase $t_a < J/4$ and the topological phase $t_a > J/4$. Dirac points $\vk_D$  are marked with red dots.}
    \label{fig_Texture}
\end{figure}

We want to ensure that a TPT from a band/trivial insulator (BI) to a topological insulator/metal (TI) phase exists when crossing the critical point $t_a^C$. For this purpose, the Berry phase $\oint_\gamma\bm{A} \cdot d\vk$ is analized for a given loop $\gamma$, with 
$$
\bm{A}^\s_\pm(\vk) = -i \langle \psi^\s_\pm(\vk)| \nabla_{\vk} | \psi^\s_\pm(\vk) \rangle
$$
the Berry potential for Hamiltonian eigenstates \eqref{eigenH} properly normalized. For $d_y^\s=0$ the system possesses PT symmetry and we have real eigenvectors $|\psi^\s_\pm(\vk) \rangle$ which give zero Berry potential and curvature, except for singularities at the nodal points where the projection is undefined; therefore, for this analysis we resort to the complex (gauge equivalent) version \eqref{gauge}. In Figure \ref{fig_Texture}(a) we represent the Berry curvature
$$B^\s_\pm(\vk)=\frac{\partial({A}^\s_\pm)_y}{\partial k_x}-\frac{\partial({A}^\s_\pm)_x}{\partial k_y} =\pm\frac{1}{2}\left(\frac{\partial \hat{\vd}_\s(\vk )}{\partial k_x}\times
\frac{\partial \hat{\vd}_\s(\vk )}{\partial k_y}\right)\cdot \hat{\vd}_\s(\vk)
$$
($\hat{\vd}_\s=\vd_\s /|\vd_\s |$) for $\s=-1$ and conduction band (+) in the insulating ($t_a<t_a^C$) and topological ($t_a>t_a^C$) phases. To numerically visualize the Dirac-delta-like divergence of the Berry curvature at the nodes, a small regularizing gap parameter ($m \approx 0.05t$) was introduced into the calculation for the plotted Figure \ref{fig_Texture}(a). This broadens the mathematical singularity into a resolvable peak without altering the quantized $\pm \pi$ Berry phase computed via contour integration.

Because the conduction and valence bands touch at isolated points in the Brillouin zone for $t_a > t_a^C$, the system is in a gapless Dirac nodal semimetal phase. In this gapless regime, a global two-dimensional Chern number is undefined. Instead, the non-trivial topology of the bulk spectrum is protected by local topological invariants: each isolated Dirac node acts as a source or sink of Berry curvature, carrying a quantized Berry phase of $\oint_\gamma\bm{A} \cdot d\vk = \pm \pi$ (or equivalently, a nodal winding number $w = \pm 1$). A transition to a fully gapped topological phase characterized by integer spin Chern numbers would require the inclusion of a bulk-gapping mechanism, such as spin-orbit coupling (SOC) or inversion-symmetry-breaking mass terms. In the absence of such terms, the system remains a gapless nodal semimetal whose edge states are protected by the bulk-boundary correspondence associated with the non-zero nodal winding numbers.

We also visualize the TPT through the winding of the pseudospin texture (the three-vector $\vd$) around the nodal points. In Figure \ref{fig_Texture}(b) we also plot the vector field  $(d_x^\s,d_y^\s)$ texture (for the real Hamiltonian) in the BI and TI phases. For $t_a<t_a^C$, $\vd$-streams flow smoothly without swirls (zero winding number). For $t_a>t_a^C$, one sees two distinct vortices centered at the Dirac (red) points. In your walk around a vortex, the vector rotates by $360^\circ$, which corresponds to a Berry phase of $\pi$ for the spinor wavefunction. We will provide a complementary (information-theoretic) characterization of the TPT in Section \ref{suscepsec} inside the real-space (versus momentum-space) picture of altermagnetism.

Let's take a closer look at the spectral structure of the AM Hamiltonian.

\section{Spin-splitting, Effective Altermagnetic Strength and Anisotropy of Conductivity.}\label{condsec}

We move from spectral properties to observable transport signatures by calculating the spin-splitting and group-velocity distributions across the Fermi surface. This allows us to quantify the giant conductivity anisotropy and spin-steering effects that distinguish altermagnets from conventional antiferromagnetic materials, and that characterize the $\dd$-wave phase, identifying the specific crystal axes along which spin-polarized currents can be most effectively steered.

\subsection{Spin splitting and effective altermagnetic strength}

\begin{figure}[h]
\begin{minipage}[b]{0.38\columnwidth}
        \includegraphics[width=\textwidth]{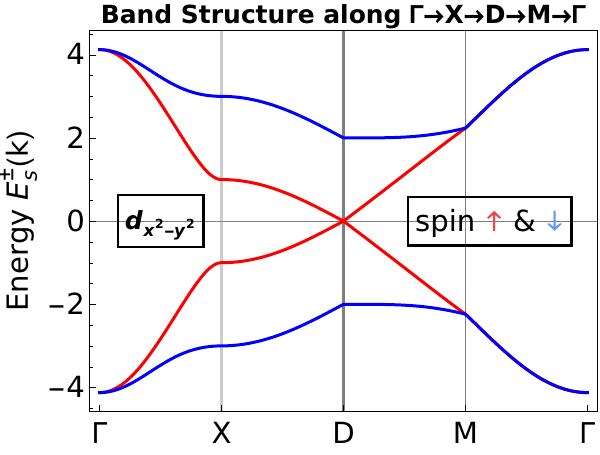}
        \centerline{(a)} 
    \end{minipage}
    \hfill
    \begin{minipage}[b]{0.52\columnwidth}
        \includegraphics[width=\textwidth]{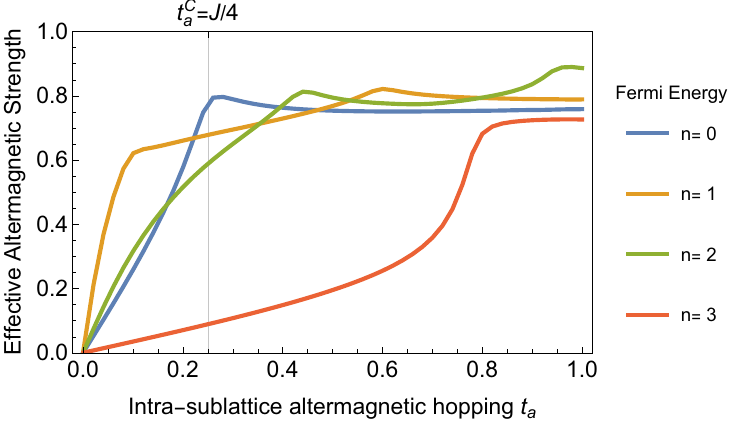}
        \centerline{(b)} 
    \end{minipage}
	\captionsetup{justification=raggedright, singlelinecheck=false}
	\caption{(a) Spin splitting along the closed trajectory $\Gamma\to X\to D\to M\to \Gamma$ in BZ for $J=1, t_a=0.5$ (in $t$ units). (b) $P_\mathrm{eff}$ versus $t_a$ for four values of the Fermi energy  $E_F(n)=n\sqrt{J^2+16t^2}/4, n=0,1,2,3$. The critical point $t_a^C=J/4$ is marked with a vertical gridline. Model parameters: $J=1$ and broadening $\eta=0.1$ (in $t$ units).
	}
	\label{fig_EffectiveAMStrength}
\end{figure}

In a conventional FM, spin polarization is a global property: you have more $\uparrow$ electrons than $\downarrow$ electrons throughout the entire crystal. In an AM, the global polarization is zero, but the local polarization in momentum space is high. The spin splitting at a fixed momentum $\vk$ is defined as follows
\be
\Delta E(\vk) = |E_{\uparrow}(\vk) - E_{\downarrow}(\vk)|.
\ee
For our specific $\dd$-wave AM, and for small $t_{a}$ and $J$, this splitting is dominated by the AM term $|d_z^\s(\vk)|$ in \eqref{dzAM}, which attains the values $|d_z^\s(0,\pi)|=|\s J+4t_a|$ and $|d_z^\s(\pi,0)|=|\s J-4t_a|$ at points $X=(\pi,0)$ and $Y=(0,\pi)$ of the FBZ. The spin splitting at those points is maximum with the value
\be
\Delta E(X)=||J-4t_a|-|J+4 t_a||=\left\{\ba{lcl}  8t_a,&\mathrm{if}& t_a<t_a^c,\\  2J,& \mathrm{if}& t_a\geq t_a^c.\ea\right.,\label{maxss}
\ee
which exibits a ``clip'', or saturated ramp, behavior as a function of $t_a$.  Spin splitting at the Dirac points is $2J$, coinciding with maximum spin splitting at point $X$ for $t_a\geq t_a^C$.

The variation of spin splitting along the closed path $\Gamma\to X\to D\to M\to \Gamma$, drawn in the FBZ of Figure \ref{fig_AMlattice}, can be appreciated from the band structure in Figure \ref{fig_BandStructure}(a) along this path.

Similarly, the local spin polarization at the Fermi level $E_F$ is defined as follows
\be
P(\vk) = \frac{A_{\uparrow}(\vk, E_F) - A_{\downarrow}(\vk, E_F)}{A_{\uparrow}(\vk, E_F) + A_{\downarrow}(\vk, E_F)}
\ee
with
\be
A_{\s}(\vk, E_F) \approx \delta_\eta(E_F - E_{\s}(\vk))
\ee
the spectral weight function (density of states), with the Dirac delta often approximated by the Lorenzian
$\delta_\eta(x)=(\eta/\pi)/(x^2+\eta^2)$ for a given broadening $\eta$ defining the Fermi Surface (FS) thickness. FS is defined by the set of points in $\vk$-space where the energy equals  $E_F$. In our numerical evaluation of the spectral functions, we employ a finite broadening factor of $\eta = 0.1t$. Assuming a typical hopping scale of $t \approx 1$ eV, this corresponds to a physical energy broadening of $\sim 100$ meV (or a quasiparticle lifetime of $\tau \approx 3.3$ fs). This choice physically accounts for realistic intrinsic scattering rates (e.g., impurities, phonons) in transition metal oxides at finite temperatures, capturing the combined effects of impurity scattering, electron-phonon interactions, and thermal smearing (where $k_B T \approx 25$ meV at room temperature). Furthermore, because our total bandwidth is roughly $4t$ and our exchange splitting is $J=1t$, a broadening of $\eta = 0.1t$ correctly preserves the macroscopic band separation while smoothly resolving the spectral weight near the nodal crossings, ensuring our polarization calculations in Eq. \eqref{eqPeff} and the results in Fig. \ref{fig_EffectiveAMStrength}(b) are physically robust and qualitatively independent of the exact choice of $\eta$, provided $\eta$ is within a physically reasonable range for standard quasiparticle lifetimes.

Because of the AM symmetry, for every point $\vk$ with positive polarization, there is a point $\vk'$ (rotated by $90^\circ$) with equal and opposite negative polarization. Therefore, the total/net polarization
\be
P_\mathrm{net} = \frac{1}{4\pi^2} \int_{BZ} P(\vk) d^2\vk
\ee
is zero. This is the reason why an altermagnet shows no macroscopic magnetization and produces no stray fields. Therefore, to measure the ``strength'' of the AM effect, we take the absolute value of the local polarization. This indicates the average degree of spin-purity of electron if we select a random direction $\vk$ in momentum space. The Effective Polarization
\be
P_\mathrm{eff} = \frac{1}{4\pi^2} \int_{BZ} |P(\vk)| d^2\vk\label{eqPeff}
\ee
quantifies how much ``usable'' spin is available for transport by measuring the magnitude of the spin-splitting $t_a$, regardless of its sign. In Figure \ref{fig_EffectiveAMStrength}(b) we represent $P_\mathrm{eff}$ as a function of the AM strength $t_a$ for four equi-spaced values of $E_F\in[0,E_\mathrm{max}]$ with $E_\mathrm{max}$ the conduction band maximum energy in \eqref{cbme}. The behavior of the dependence of $P_\mathrm{eff}$ on $t_a$ resembles that of the saturated ramp giving the maximum spin splitting $\Delta E(X)$ in \eqref{maxss}. For $t_a<J/4$ (band insulator phase), the effective polarization is an increasing function of $t_a$. For $t_a>J/4$ (topological insulator phase), the effective polarization attains values near $0.8$, except for Fermi energies close to $E_\mathrm{max}$ where the density of states is lower.

$P_\mathrm{eff}$ matters for technology.  A value of $P_\mathrm{eff} = 0.8$ means that, on average, an electron at the FS is $80\%$ ``spin-pure''. In the next section we will see that AM materials come with a privileged direction for electric conduction. Because polarization is tied to $\vk$, we can ``filter'' electrons. If we created a device that only allowed transport along the maximum spin-splitting (nodal) directions $X$ and $Y$ (or the diagonals for the $\dd_{xy}$ or [110] alignement), the operational polarization would approach $100\%$, even though the material is globally ``antiferromagnetic''.

Next we analyze the group velocity distribution across the FS and giant AM conductivity anisotropy.

\subsection{Group velocity distribution across the Fermi surface and anisotropy of conductivity}

   \begin{figure}[h]
	\begin{center}
		\includegraphics[width=0.35\columnwidth]{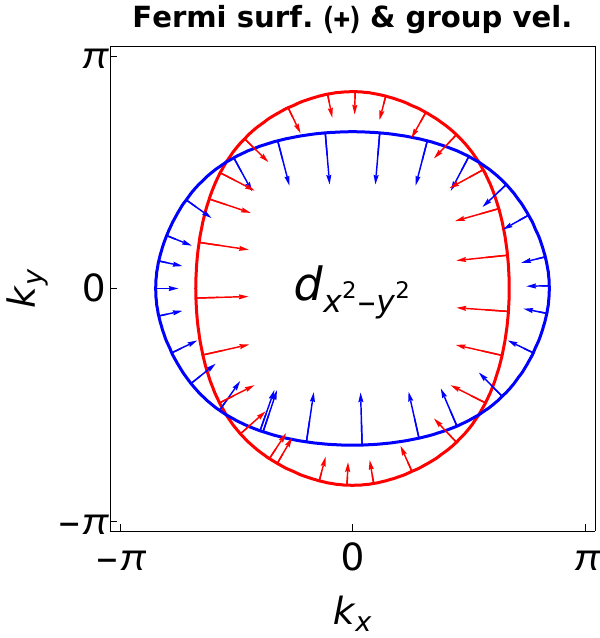}
		\includegraphics[width=0.35\columnwidth]{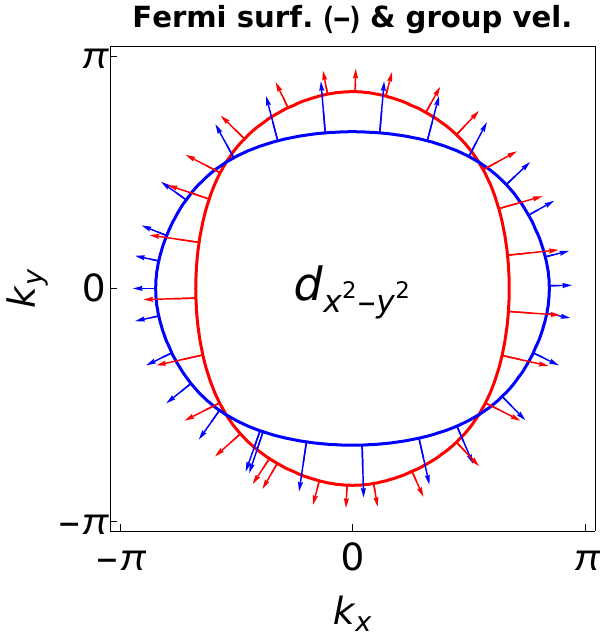}
	\end{center}
	\captionsetup{justification=raggedright, singlelinecheck=false}
	\caption{Group velocity distribution across the Fermi surface ($E_F=\pm 2$), for conduction $(+)$ and valence $(-)$ bands  and spin up (red) and down (blue).  Model parameters: $J=1, t_a=0.2$ ($t$ units).
	}
	\label{fig_FermiSurfacedx2my2}
\end{figure}

To understand why AMs generate spin currents, we have to look at the group velocity
\be
\bm{v}_\s(\vk) = \frac{1}{\hbar} \nabla_{\vk} E_{\s}(\vk).
\ee
This vector indicates the direction and speed at which an electron with momentum $\vk$ and spin $\s$ actually moves through the crystal.

To visualize which spin dominates transport, we look at the FS. In an AM, the FS for spin-up and spin-down electrons  will have different shapes, even though they enclose the same total area. This ``shape anisotropy''  is why  spin-polarized effects can be obtained without a net magnetic moment.

In Figure \ref{fig_FermiSurfacedx2my2}, we draw $E_{\s}^\pm(\vk)=E_F^\pm=\pm 2$ FS contours for $J=1, t_a=0.2$ (in $t$ units). FSs appear as two concentric-like loops, but with a twist for spin up (red) and down (blue), for valence (-) and conduction (+) electrons. The blue loops are stretched along the $X$ direction, while the red loops  are  stretched along the $Y$ direction. The two loops cross at the diagonals $k_x=\pm k_y$ of the BZ, where the AM term $(\cos k_x -\cos k_y)$ vanishes, causing spin degeneration. This visualization explains the anomalous Hall effect in AMs \cite{PhysRevB.110.094425}. Let us analyze the transport implications.

In Figure \ref{fig_FermiSurfacedx2my2} we also represent the group velocity $\bm{v}_\s(\vk)$ (in arbitrary units) for spin-up (red arrows) and spin-down (blue arrows) valence (-) and conduction (+) electrons at the FS.  The group velocities turn out to be highly directional (anisotropic), displaying a ``steering'' effect.
Since the electron velocity $\bm{v}_\s(\vk)$ is the slope of $E_{\s}(\vk)$,
the preference comes from how the altermagnetic energy term modifies the electron's speed (group velocity) in different directions. For spin $\uparrow$, the altermagnetic term adds to the kinetic energy slope along the $X$ direction,  the effective mass drops, and the electron becomes fast. Along $Y$, the AM term subtracts from the slope, the effective mass increases and the electron becomes slow. For spin $\downarrow$ the signs flip: electrons are slow along $X$ and fast along $Y$. For [110] aligned $\dd_{xy}$-wave AMs, the fast nodal lines are the diagonals: $X=Y$ for spin $\uparrow$ and $X=-Y$ for spin $\downarrow$. This AM effect creates a spin-polarized current (longitudinal), rather than a spin-Hall current (transverse).

Therefore, in a $\dd_{x^2-y^2}$-wave AM, the ``steering'' of electrons is spin-dependent. Even if an electric field is applied along the diagonals, the electrons are ``pushed'' toward the axes $X$ and $Y$ because of the warped shape of the energy bands. Consequently, electrical resistance will depend heavily on the direction of the current. See Figure \ref{fig_GapFET}(b) for a later FET proposal based in this AM feature.

\begin{figure}[h]
	\begin{center}
	\includegraphics[width=0.31\columnwidth]{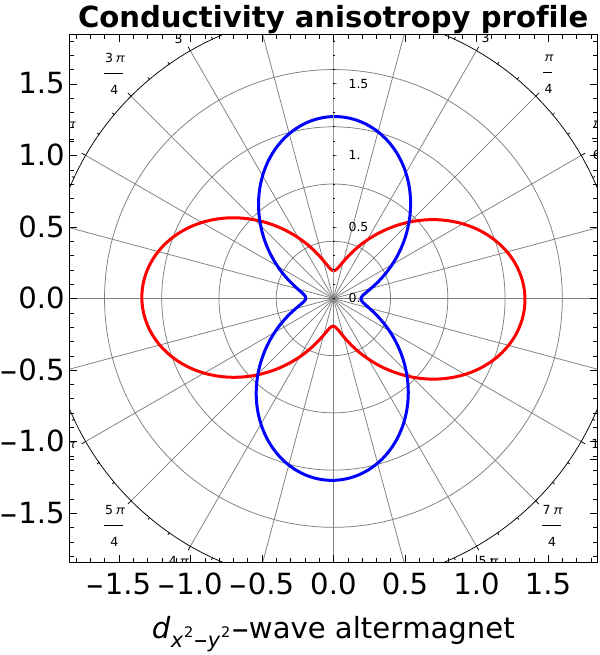}
	\end{center}
	\captionsetup{justification=raggedright, singlelinecheck=false}
	\caption{Polar plot showing the conductivity's angular dependence for the  $\dd_{x^2-y^2}$-wave AM. Red and blue represent the spin-resolved conductivities, $\sigma^\uparrow$ and $\sigma^\downarrow$, respectively. The anisotropic lobes correspond to the maximum group velocity directions. Model parameters: $J=1, t_a=0.2$ (in $t$ units).
	}
	\label{fig_SpinConductivityTensor}
\end{figure}

Next we explain the implications of AM on conductivity anisotropy profiles.
We compute the Drude weight $\mathcal{D}$ (at zero temperature) by integrating the squared group velocity over the Fermi surface (FS)
\begin{equation}
\mathcal{D}_{ij}^\s= \int_{\text{BZ}}\frac{d^2\vk}{(2\pi)^2} v^i_\s(\vk)v^j_\s(\vk)\delta(E_\s(\vk)-E_F)=\oint_{\text{FS}^\s} \frac{d\ell(\vk)}{(2\pi)^2\hbar} \, \frac{v^i_\s(\vk) \, v^j_\s(\vk)}{ |\bm{v}_\s(\vk)|},\; i,j=x, y,
\end{equation}
where ${d\ell}/(\hbar|\bm{v}|)$ is the local density of states for a specific segment $d\ell$ of the Fermi surface. It arises as the Jacobian in the coordinate transformation $d^2\vk = dk_\perp \, d\ell = dE\, d\ell/(\hbar|\bm{v}|)$, when you convert the integral over the full 2D area of the BZ into a 1D line integral along the FS. This quantity characterizes the coherent charge transport and allows us to estimate the conductivity $\sigma_{ij}^s = e^2 \tau \mathcal{D}_{ij}^s$,  with $e$ the electron charge and $\tau$ the scattering time. We will ignore the units and refer to $\mathcal{D}^\s$ simply as the spin conductivity tensor $\vs^\s$ measuring the relationship between an applied electric field and the resulting spin-polarized flow.

In Figure \ref{fig_SpinConductivityTensor} we show the conductivity anisotropy profile as a polar plot of $\sigma^\s(\theta)=\bm{n}^T\vs^\s\bm{n}$ as a function of the unit vector $\bm{n}=(\cos\theta,\sin\theta)$ polar angle $\theta$. Red and blue represent the spin-resolved conductivities, $\sigma^\uparrow$ and $\sigma^\downarrow$. Fast electrons (high group velocity, see Figure \ref{fig_FermiSurfacedx2my2}) dictate the direction of maximum conductivity.  The $C_4 \mathcal{T}$ symmetry is indeed perfectly imprinted on the results: applying a $C_4$ rotation ($\theta \to \theta + 90^\circ$) and the time-reversal operator $\mathcal{T}$ (which flips spin, $\uparrow \leftrightarrow \downarrow$) leaves the system invariant. This is directly visible in the plot, where the red curve is exactly mapped onto the blue curve upon a $90^\circ$ shift. This is a beautiful manifestation of the altermagnetic symmetry in the transport coefficients.

While $\dd_{x^2-y^2}$-wave AM [100] aligned  materials conduct current better along the principal axes, $\dd_{xy}$-wave AM  [110] aligned materials conduct current well along diagonals.  A transverse spin current is generated without any external magnetic field or relativistic spin-orbit coupling. We shall explore this fact for a FET proposal in AM  nanoribbons in Section \ref{gapsec}.

The quantity $\mathcal{Q}=\sigma^{\uparrow} - \sigma^{\downarrow}$ represents the spin conductivity tensor. It is an important quantity of AM spintronic to quantify how much net spin is transported when we apply an electric field $\bm{E}$. The electric current for each spin channel is $\bm{J}_s=\vs^\s\bm{E}$. The charge current (measured with an ammeter) is the sum $\bm{J}_\mathrm{charge}=\bm{J}_\uparrow+\bm{J}_\downarrow$, whereas the  spin current (the flow of angular momentum) is the difference $\bm{J}_\mathrm{spin}=\bm{J}_\uparrow-\bm{J}_\downarrow$. Therefore, the conductivity tensor $\mathcal{Q}$ determines the material's ability to generate a spin current from a charge voltage. In a standard AFM, the two sublattices $A$ and $B$ are identical but inverted, so that symmetry dictates that $\mathcal{Q} = 0$; that is, both spins move at the same speed when we apply a voltage (there is no net flow of spin). However, in an AM material, although the magnetization remains zero as in the case of AFM, yet $\vs^{\uparrow} \neq \vs^{\downarrow}$ in specific directions due to the group velocity mismatch. Consequently, if we construct a device using a $\dd_{x^2-y^2}$-wave AM and apply an electric field along the $X$-axis, we measure a current but, since spin up electrons are moving faster than spin down electrons, the current exiting the material is highly spin-polarized. As we have already commented in the Introduction, this also implies a SST
that makes AM highly desirable as an alternative to more traditional  SOT-MRAM  (see \cite{Nguyen2024} for recent progress), where a pure spin current is generated to flip a magnetic bit, without the need for the stray magnetic fields of a FM.

\section{Real-Space Nanoribbon Structure Analysis}\label{realspacesec}

We now turn into a real-space representation of our AM by performing an inverse Fourier transform. We study a 2D sample (a nanoribbon) of area $A=L_xL_y$ (in $a^2$ units), so that our spin-up and spin-down Hamiltonian matrices are both of size $D=2A$.

As an alternative to momentum-space invariants, we introduce the fidelity-susceptibility and IPR to diagnose the topological phase transition in real space. This information-theoretic markers are particularly robust for characterizing finite systems, such as narrow nanoribbons, where the closing of the bulk gap and the emergence of edge states can be detected without relying on translational symmetry.

\subsection{Fidelity-susceptibility and IPR}\label{suscepsec}

\begin{figure}[h]
	\begin{center}
		\includegraphics[width=0.45\columnwidth]{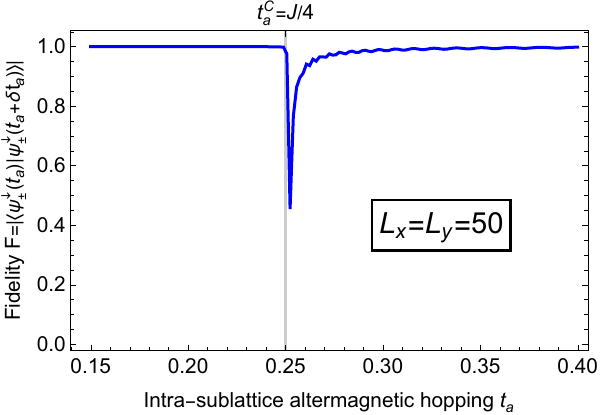}
		\includegraphics[width=0.45\columnwidth]{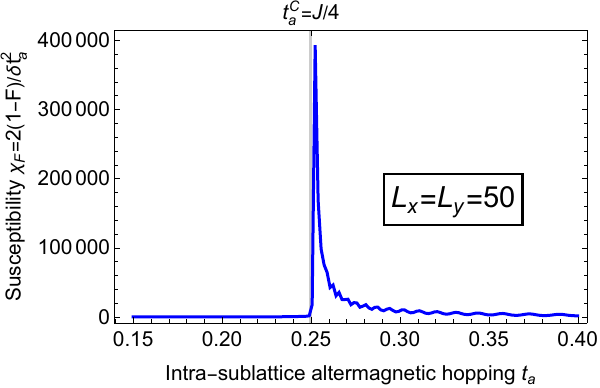}\\
	\end{center}
	\captionsetup{justification=raggedright, singlelinecheck=false}
	\caption{Fidelity and fidelity-susceptibility of spin-down edge states (conduction and valence) as a function of $t_a$  for a sample of size $L_x = L_y = 50$ (same result for spin-up). The TPT critical point $t_a^C$ is marked with a vertical grid line. Model parameters: $J = 1\rightarrow t_a^C=J/4=0.25$ ($t$ and $a$ units).
	}
	\label{fig_Fidelity}
\end{figure}

The characterization of topological phases is usually done in momentum space through the Berry phase and Chern, winding, Pontryagin, etc., topological invariants, as we have seen in Section \ref{berrysec}. TPTs fall outside the Landau framework of QPTs (usually characterized by a spontaneous symmetry breaking) because the two topological phases can have exactly the same symmetry, although different topology. In addition, while QPTs are characterized by local order parameters (e.g., magnetization), for TPTs we have global topological invariants. However, QPTs and TPTs still share similarities. The effective AM strength represented in Figure \ref{fig_EffectiveAMStrength}(b) can be still seen as a kind of order parameter with lower values in the BI phase and higher  values in the TI phase. Also, typical information theory markers of QPTs [like fidelity-susceptibility, entanglement entropy, delocalization measures, etc.] have also proven to be useful for characterizing TPTs; see e.g. \cite{silicene1,silicene2,silicene3} for Xenes, \cite{IJQC,MRX} for phosphorene, \cite{CALIXTO2022128057,SciPostPhys.16.3.077,IJMPB2025} for HgTe/CdTe quantum wells and \cite{TBGPhysicaE} for twisted bilayer graphene. Let us see how fidelity-susceptibility characterizes our  TPT in AM across the critical point $t_a^C$.

Just as the occurrence of QPTs is often marked by a sudden alteration in the structure of the ground-state wavefunction, the phenomenon of TPTs is also accompanied by an abrupt change in the structure of edge states $\psi(t_a)$ at $t_a=t_a^C$. This change is captured by the fidelity \cite{Gu,Zanardi}
\be
F(t_a,\delta t_a)=|\langle \psi(t_a)|\psi(t_a+\delta t_a)\rangle|,
\ee
which measures the similarity between two close states. Whereas fidelity $F(t_a,\delta t_a)$ is highly sensitive to the particular choice of $\delta t_a$, its leading (quadratic) order change $F(t_a,\delta t_a)\approx 1-\frac{1}{2}\chi_F(t_a)(\delta t_a)^2+O(\delta t_a^3)$, that is, the fidelity-susceptibility 
\be \chi_F(t_a)=-\left.\frac{\partial^2F(t_a,\delta t_a)}{\partial (\delta t_a)^2}\right|_{\delta t_a=0}\approx 2\frac{1-F(t_a,\delta t_a)}{\delta t_a^2}, \quad \delta t_a\ll 1,
\ee
has the advantage of reducing the  dependence on the practical choice of the probe parameter $\delta t_a$ with respect to $F$. This fact makes $\chi_F(x)$ an incredibly robust tool for detecting quantum phase transitions regardless of the control parameter $x$ grid precision. The susceptibility $\chi_F$ is also related to quantum Fisher  information by $\mathcal{F}=4\chi_F$ and therefore to sensing precision to estimate $t_a$ though the quantum Cramer-Rao bound $\Delta t_a=1/\sqrt{n \mathcal{F}(t_a)}$, with $n$ the number of measurement repetitions; this bound dictates the best possible precision allowed by quantum mechanics for estimating a parameter.

In Figure \ref{fig_Fidelity} we represent $F$ and $\chi_F$ for edge (near gap) states (either valence or conduction) as a function of $t_a$ in a neighbourhood of the critical point $t_a^C=J/4$ for $L_x=L_y=50$ and $J = 1$ (in $t$ and $a$ units). An abrupt drop of fidelity and a snappish peak in susceptibility are clearly seen at $t_a^C$. Therefore, the sensing precision is highly enhanced at $t_a^C$. This provides an information-theoretic characterization of the TPT that is complementary to the Berry picture, being also suitable for finite systems.

Here, a comment is in order. While $t_a$ serves as our theoretical control parameter to map the zero-temperature quantum phase transitions [relics of which can be probed experimentally via strain engineering or in ultracold atom systems] an equivalent transition can be traversed in a fixed material by thermally tuning the exchange field $J(T)$ below the magnetic ordering temperature, as the critical boundary depends explicitly on the dimensionless ratio $t_a/J$ [that is, the phase transition is purely a function of the competition between the kinetic energy (hopping $t_a$) and the magnetic exchange splitting ($J$)]. Nevertheless, we will continue to use $t_a$ instead of $J$ as the control parameter.

The primary advantage of real-space markers like fidelity-susceptibility is that they do not require translational invariance or a well-defined $k$-space, unlike the Chern number. This makes them exceptionally powerful for identifying topological phases in finite-size device geometries, disordered nanoribbons, or systems with impurities. In Fig. \ref{fig_Fidelity}, the fidelity-susceptibility acts as a highly sensitive detector for the Topological Phase Transition (identifying exactly where the gap closes). 

\begin{figure}[H]
	\begin{center}
		\includegraphics[width=0.45\columnwidth]{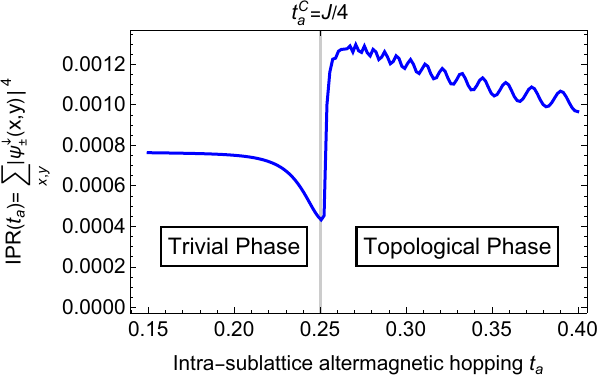}
	\end{center}
	\captionsetup{justification=raggedright, singlelinecheck=false}
	\caption{Inverse participation ratio for spin-down edge states (conduction and valence) as a function of $t_a$   for a sample of size $L_x = L_y = 50$. The critical point $t_a^C$ separating trival from topological phases is marked with a vertical grid line. Model parameters: $J = 1\rightarrow t_a^C=J/4=0.25$ ($t$ and $a$ units).
	}
	\label{fig_IPRpos}
\end{figure}

To better distinguish the trivial from the topological phases, we can use the Inverse Participation Ratio (IPR). IPR has already been used in Reference \cite{silicene3} to distinguish the trivial from the topological phase in silicene under perpendicular electric and magnetic fields, giving a sharp and clear marker. In general, IPR  measures the spreading of a vector's expansion $|\psi\ra=\sum_{n=1}^N c_n|n\ra$ in a given basis $\{|n\ra, n=1,\dots,N\}$. For a normalized vector $|\psi\ra$, the probabilities  $p_n=|c_n|^2$ to find $|\psi\ra$ in  $|n\ra$ fulfill $\sum_{n=1}^N p_n=1$. The inverse participation ratio of $\psi$ is then defined as follows
\begin{equation}
 \mathrm{IPR}_\psi=\sum_{n=1}^N p_n^2=\sum_{n=1}^N |c_n|^4.\label{IPReq}
\end{equation}
The smallest value  $\mathrm{IPR}_\psi=1/N$ corresponds to an equally weighted superposition $|c_n|=1/\sqrt{N}, \forall n$ (fully delocalized state), whereas the highest value  $\mathrm{IPR}_\psi=1$ corresponds to a perfectly localized state, that is, $c_n=\delta_{n,n_0}$ for some $n_0$.  The concept can be similarly extended to the representation of the wavevectors in position, momentum, or other spaces,  to study the vector's  spreading or localization in the given space basis. For edge states $\psi^s_\pm(x,y)$ in a AM  nanoribbon of size $L_xL_y$, the IPR can be defined as
\begin{equation}
 \mathrm{IPR}_{\psi^s_\pm}=\sum_{x=1}^{L_x}\sum_{y=1}^{L_y}|\psi^s_\pm(x,y)|^4.
\end{equation}
For reference, a fully delocalized (bulk-type) state in a sample of $L_x=50=L_y$ has IPR=$1/(2L_xL_y)=0.0002$. In Figure \ref{fig_IPRpos} we represent the IPR for a spin-down (conduction or valence) edge state $\psi^\downarrow_\pm(x,y)$ as a function of the AM strength $t_a$. The IPR immediately informs us whether the states emerging after the transition are topologically protected edge states (high localization / high IPR) or trivial bulk states (low IPR). Therefore, the primary advantage of real-space markers like fidelity-susceptibility and IPR is that they substitute the need for $k$-space invariants in finite devices.

Actually, due to finite-size quantum confinement within the narrow ribbon geometry, the baseline IPR in the trivial phase is inherently higher than the ideal infinite bulk value. Consequently, the operational criterion for diagnosing the topological phase transition is not the absolute value of the IPR, but rather the sharp discontinuity in its derivative at $t_a = 0.25$, which marks the abrupt manifestation of exponentially localized edge modes. Therefore, we could propose any of the ``local monotonicity indicators''
\[ C_F(t_a)=\sgn(F'(t_a)),\quad C_{\chi_F}(t_a)=\sgn(\chi_F'(t_a)), \quad C_{\mathrm{IPR}}(t_a)=\sgn(\mathrm{IPR}'(t_a)),\]
in a small neighbourhood of $t_a=t_a^C$ as an alternative to topological numbers to distinguish our topological phases.

\subsection{Spatial texture and chirality}

\begin{figure}[h]
	\begin{center}
		\includegraphics[width=1\columnwidth]{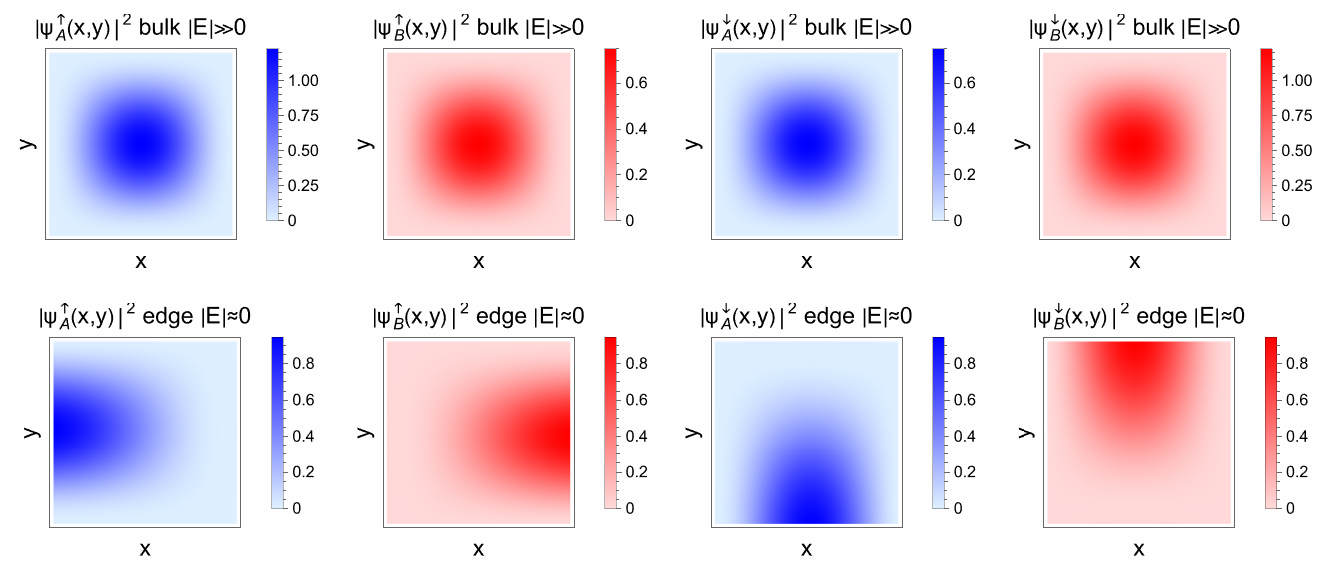}\\
	\end{center}
	\captionsetup{justification=raggedright, singlelinecheck=false}
	\caption{Spinor probability density $(|\psi_A^\uparrow|^2,|\psi_B^\uparrow|^2,|\psi_A^\downarrow|^2,|\psi_B^\downarrow|^2)_\pm$ as a function of $(x,y)\in [1,100]\times [1,100]$ for bulk (top row) and edge (bottom row) Hamiltonian eigenstates. Densities are renormalized by $L_xL_y$ to range from 0 to $\sim 1$. Model  parameters:  $J=1, t_a=J/4=t_a^C$ ($t$ and $a$ units).
	}
	\label{fig_DensityBulkEdge}
\end{figure}

Let us now look at the real-space Hamiltonian wavefunctions segregation in a finite AM cluster. We will focus on bulk (highest energy level) and edge [energy floor for conduction ``LUMO'' and energy ceiling for valence ``HOMO'') states. 

For spin $s$, the real-space Hamiltonian matrix $H_s$ has size $D=2L_xL_y$. The Hamiltonian eigenstates $\psi^s_\tau(x,y)$ are normalized according to 
$$\sum_{\tau=A, B}\sum_{x=1}^{L_x}\sum_{y=1}^{L_y}|\psi^s_\tau(x,y)|^2=1.$$

In Figure \ref{fig_DensityBulkEdge} we represent the spinor probability density for a sample of size $L_x=100=L_y$. We  separate for the two components $(\psi^\s_A,\psi^\s_B)$, for spin up and down electrons. We have chosen the valence band minimum and conduction band maximum (with identical densities) energies \eqref{cbme}  (referred to as bulk states $|E|\gg 0$, top row in Figure \ref{fig_DensityBulkEdge}) and the HOMO and LUMO energies (referred to as edge states $|E|\sim 0$, bottom row in Figure \ref{fig_DensityBulkEdge}). To get an idea of the order of magnitude, a completely delocalized state would have a probability density of $|\psi^s_\tau(x,y)|^2=1/D=0.00005$. For aesthetic reasons, to avoid such small numbers in the color bar that reduce the width of the density plots, the probability amplitudes have been rescaled by $\sim D$, more specifically $|\psi^s_\tau(x,y)|^2\to D|\psi^s_\tau(x,y)|^2/2$, so that renormalized densities now scale from 0 to $\gtrsim 1$ in the graphs. 

In the momentum-space band structure, spin polarization is  observed through energy splitting $E_{\uparrow}(\vk) \neq E_{\downarrow}(\vk)$ (in general). However, in real space, spin polarization becomes a spin density texture or chirality near boundaries or defects. The total set of energy levels is identical $\{E_{\uparrow}\} = \{E_{\downarrow}\}$ but the wavefunctions are different: $|\psi^{\uparrow}(x,y)|^2 \neq |\psi^{\downarrow}(x,y)|^2$.

Bulk electrons are localized toward the center $(x_c,y_c)=(L_x/2,L_y/2)$ of the sample, the probability distributions $|\psi^\uparrow_{\tau}(x,y)|^2$ being sharper (more concentrated) at $(x_c,y_c)$  in sublatice $\tau=A$ (blue density plots) than in sublatice $\tau=B$ (red density plots); the opposite occurs for spin $\downarrow$. In contrast,  edge states are more concentrated at the sample boundaries.  Since we know that spin up (down) electrons prefer to move along the $X$ ($Y$) direction, they will ``pile up'' differently when they hit the walls of the finite sample, resulting in a spin quadrupole pattern in real space. Indeed, spin $\uparrow$ electrons pile up to the west of the sample at sublattice $A$ and to the east at sublatice $B$, whereas spin $\downarrow$ electrons accumulate to the south at sublattice $A$ and to the north at sublatice $B$. For $\dd_{xy}$-wave AMs, probability density concentrates along diagonals, $x=y$ for spin up and $x=-y$ for spin down. Note that if the plot is rotated by $90^\circ$, red becomes blue, reflecting the $\dd$-wave symmetry.

\subsection{Global and local density of states}\label{LDOSsec}

\begin{figure}[h]
	\begin{center}
		\includegraphics[width=0.5\columnwidth]{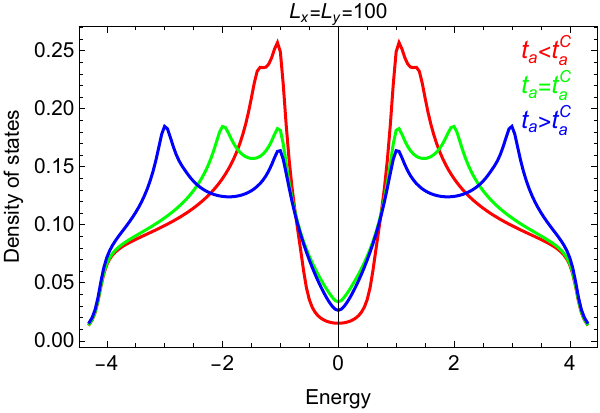}
	\end{center}
	\captionsetup{justification=raggedright, singlelinecheck=false}
	\caption{DOS for spin up (similar results for spin down) in the BI  ($t_a=0.1$, left) and TI ($t_a=0.5$, right) phases, and critical $t_a=0.25$, for a sample of $L_x=L_y=100$.  Model parameters: $J=1$, $\eta=0.1$ ($t$ and $a$ units).
	}
	\label{fig_DOS}
\end{figure}

A common characteristic of topological phases is the developement of topological zero-energy (flat bands) edge states. This is an important phenomenon related to the appearance of van Hove singularities at zero energy, which contribute to the sharp peaks
in the corresponding density of states (DOS) measured as follows
\be
\mathrm{DOS}_\eta(E)=\frac{1}{D}\sum_{n=1}^D \delta_\eta(E-E_n)=\frac{1}{D} \sum_{n=1}^D  \frac{\eta/\pi}{(E-E_n)^2 + \eta^2},
\ee
where $E_n$  are the Hamiltonian eigenvalues and $D=2L_xL_y$ is the total number of Hamiltonian $H_s$ eigenstates. Low-energy flat bands also develop at magic angles in twisted bilayer graphene 
\cite{Bistritzer12233}. This flattening of moir\'e bands at magic angles is accompanied by a vanishing of the effective Dirac-point Fermi velocity and strong enhancement of the counterflow conductivity, as experimentally measured by 
\cite{Cao2018,Cao2018unconventional}. Flat bands and magic angles are also captured by fidelity-susceptibility and other information entropy measures \cite{TBGPhysicaE}.

In Figure \ref{fig_DOS} we show the DOS for spin-up electrons (identical for spin-down electrons) below, above and at the critical point $t_a^C$. In this case, we cannot say with absolute certainty that flat bands (near zero-energy) develop above $t_a^C$ (the peaks are not quite sharp). The DOS also does not provide information on the spin-splitting effect, because up and down electron contributions are identical. Therefore, we resort to a more convenient/interesting magnitude, such as the Local Density of States (LDOS), which combines Hamiltonian eigenvalues and eigenvectors.

\begin{figure}[h]
\begin{minipage}[b]{0.48\columnwidth}
        \includegraphics[width=\textwidth]{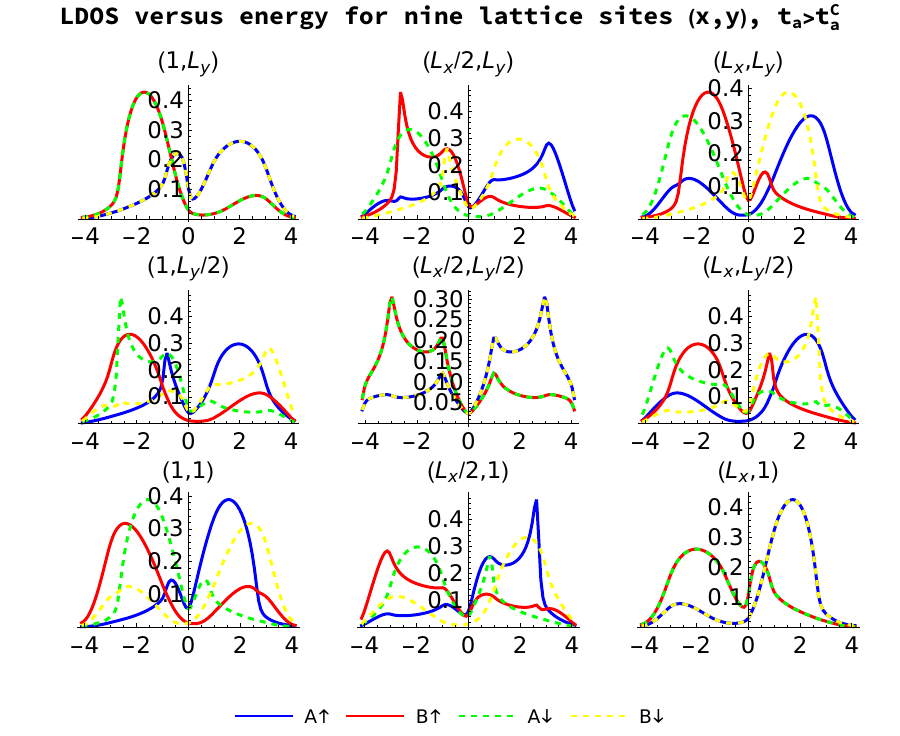}
        \centerline{(a)} 
    \end{minipage}
    \hfill
    \begin{minipage}[b]{0.48\columnwidth}
        \includegraphics[width=\textwidth]{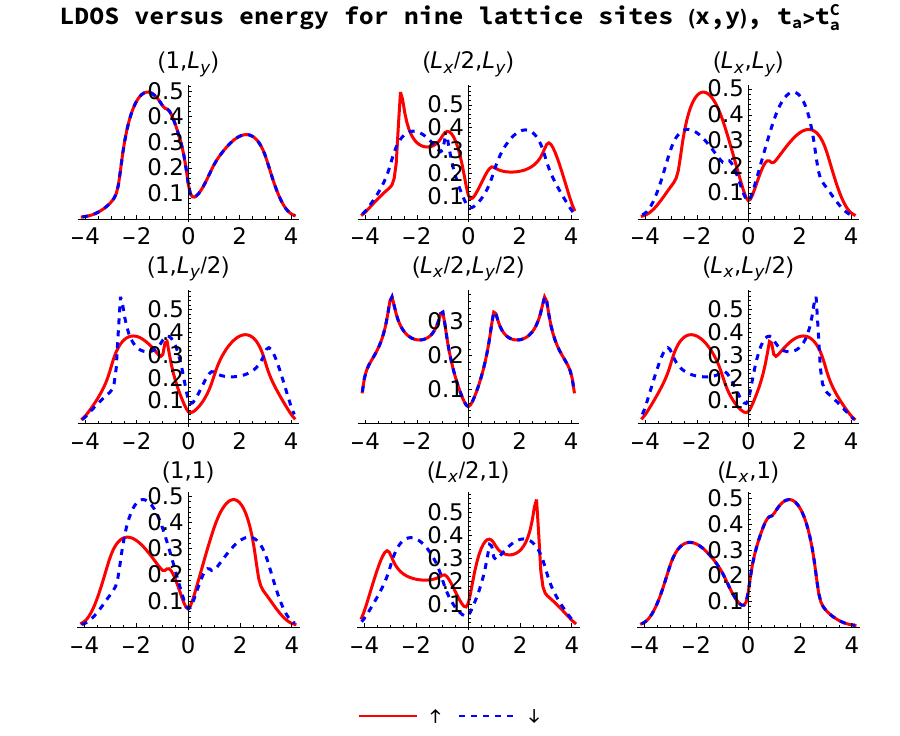}
        \centerline{(b)} 
    \end{minipage}
	\captionsetup{justification=raggedright, singlelinecheck=false}
	\caption{LDOS as a function of the energy $E$ for nine lattice sites $(x,y)$ of our rectangular lattice: corners, midpoints sides and center. (a) Resolved by sublattice and spin. (b) Resolved by spin. Sample of $L_x=L_y=100$.  Model parameters: $J=1$, $t_a=0.5$ ($t$ and $a$ units).
	}
	\label{fig_LDOS}
\end{figure}

To compute the LDOS, we evaluate how much of the total electronic density  at a specific energy $E$ is concentrated on a specific lattice site. The formula of LDOS at energy $E$ on a lattice site $(x,y)$ is
\be
\mathrm{LDOS}_\eta^{(x,y)}(E) = \sum_{n=1}^{D} |\psi_n(x,y)|^2 \delta_\eta(E - E_n)
\ee
where $E_n$ and $\psi_n$ are the eigenvalues and eigenvectors of our $D=4L_xL_y$ size (in lattice constant $a=1$ units) tight-binding matrix Hamiltonian. We want to separate the contributions to LDOS by sublattice  $\tau=A,B$ and spin $\s=\pm 1$. In Figure \ref{fig_LDOS} we plot the LDOS as a function of the energy $E$ for nine lattice sites $(x,y)$ of our rectangular lattice: corners, midpoints sides and center. The (a) panel is resolved by sublattice and spin, whereas the (b) panel is resolved just by spin (i.e., the combined  $A+B$ contributions).

Based on both LDOS plots provided, we draw some conclusions about our sample's physics. LDOS distinguishes lattice, spin, valence and conduction contributions at different edges and vertices. In general, the electronic density is more concentrated about the center of the valence and conduction bands, with a clear bias between the LDOS valence distribution ($E<0$) and the conduction distribution ($E>0$) at the different edges, but not at the center $(L_x/2, L_y/2)$ of the sample, where both contributions are balanced. This means that the bulk states are spin-degenerate (or compensated); i.e., there is no net magnetization. Disregarding sublattice $A,B$ (b  panel), we can say that an electron moving in the bulk has an equal probability of being spin-up or spin-down (note that the red and blue curves overlap perfectly). This overlap also happens for electrons located about the anti-diagonal corners, but now there is a noticeable bias between valence and conduction. This corresponds to the PT symmetry, $(x,y)\to -(x,y)$ and $E\to -E$, of our AM.

Note also that the spin-up conduction (valence) electronic density is more (less) concentrated downwards than upwards. The spin-down conduction (valence) electronic density is more (less) concentrated toward the right than toward the left. This picture is compatible with the conductivity profile of Figure \ref{fig_SpinConductivityTensor}. In fact, using a traffic analogy, $X$-axis is a highway for spin-up electrons, whereas they experience a trafic jam in the $Y$-direction. This creates piles of spin-up electrons at the bottom (for $E>0$) or at the top (for $E<0$). A similar reasoning applies to spin-down electrons applying a $90^\circ$ degrees rotation. We next discuss a split-drain FET architecture based in this AM singular characteristic.

\section{Hybridization gap  and FET proposal for nanoribbons}\label{gapsec}

Leveraging the controllable energy gap that emerges from edge-state hybridization in narrow geometries, we propose a concrete altermagnetic field-effect transistor (FET). This section demonstrates how the intrinsic group-velocity anisotropy can be harnessed for ``edgetronics'', enabling the electrostatic gating of spatially spin-polarized ballistic transport.

\begin{figure}[h]
	\begin{minipage}[b]{0.38\columnwidth}
        \includegraphics[width=\textwidth]{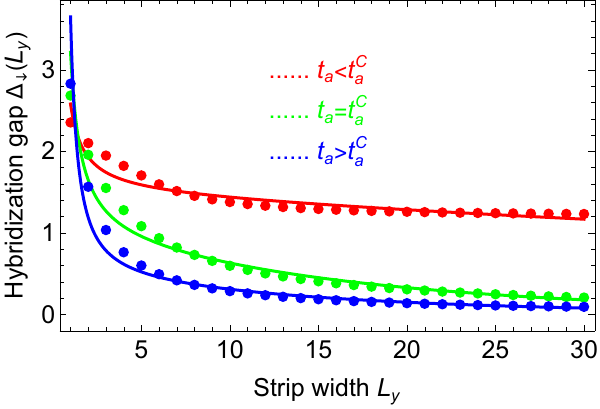}
        \centerline{(a)} 
    \end{minipage}
    \hfill
    \begin{minipage}[b]{0.52\columnwidth}
        \includegraphics[width=\textwidth]{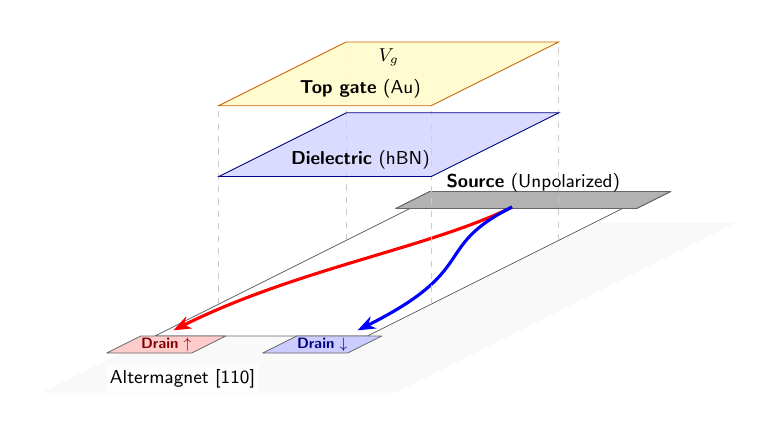}
        \centerline{(b)} 
    \end{minipage}
    \captionsetup{justification=raggedright, singlelinecheck=false}
	\caption{(a) Hybridization gap for spin-down electrons in a strip geometry of length $L_x=30$ as a function the strip width $L_y$, before, after and at the critical point $t_a^C$. Model parameters: $J = 1.0\Rightarrow t_a^C=0.25$ ($t$ and $a$ units). (b) Split-drain AM FET geometry.
	}
	\label{fig_GapFET}
\end{figure}

Let us consider nanoribbons of size $L_xL_y$ with $L_x\gg L_y$.
Edge states in these nanoribbons usually display an exponential behavior, $\psi(y)\sim\exp(-\kappa y)$,  to  account for exponential localization at the boundaries $y=0,L_y$ (see e.g. \cite{PhysRevLett.101.246807,IJMPB2025} for the case of HgTe quantum wells in an infinite strip of width $L$). When the coherence length $\lambda=1/\mathrm{Re}(\kappa)$ of edge states is comparable to the strip width $L_y$, states from opposite edges can overlap (tunneling between edges occurs), leading to a gap opening in the edge state spectrum (different from a bulk Dirac mass gap) that usually scales exponentially with the strip width, for large enough $L_y$. We propose a fitting for the hybridization gap of the following form
\be
\Delta(L_y)=\Delta_0 e^{-L_y/\lambda+\xi/L_y},\quad L_y\geq 1,\label{hgap}
\ee
which accounts for values of $L_y$ not necessarily large. In Figure \ref{fig_GapFET}(a) we plot $\Delta(L_y)$  as a function of the strip width $L_y$, for a fixed strip length of $L_x=30$, and for three values of $t_a$: below, above and at the critical point $t_a^C$. We restrict ourselves to spin down electrons; the gap for spin-up electrons exhibits a similar behavior except for low $L_y$. The values of $\Delta_0$, the coherence length $\lambda$ and the low $L_y$ behavior captured by $\xi$ are given in Table \ref{tabla}. We also include the gap  for the square sample ($L_y=L_x$) in the last column. In general, the gap decreases with the ribbon width $L_y$, more drastically for narrow  than for wide ribbons. In the topological phase $t_a\geq t_a^C$, the coherence length is $\lambda\simeq 16$ (in $a$ units). We see that, in fact, for values of $L_y$ below $\lambda$, the gap is widening significantly, reaching characteristic values of $\Delta\sim 1$ (in $t$ units) for ultra-narrow ribbons, likely $L_y < 10$, depending on the material's  parameters (remember that we are using  $J=1$ in $t$ units).

\begin{table}[h]
\begin{center}
\begin{tabular}{l|c|c|c|c|}
$\Delta(L_y)$  & \multicolumn{1}{l|}{$\Delta_0$} & \multicolumn{1}{l|}{\;$\lambda$} &  \multicolumn{1}{l|}{\;$\xi$} &\multicolumn{1}{l|}{$\Delta(30)$}    \\ \hline
\multicolumn{1}{|l|}{$t_a=0.1$} & $1.48$ & $117$ & $0.56$ & $1.23$ \\ \hline
\multicolumn{1}{|l|}{$t_a=0.25$} & $1.01$ & $17$ & $1.22$& $0.21$ \\ \hline
\multicolumn{1}{|l|}{$t_a=0.5$} & $0.47$ & $16$ & $2.12$& $0.09$ \\ \hline
\end{tabular}
 \end{center}
 \captionsetup{justification=raggedright, singlelinecheck=false}
\caption{Fitting parameter values of the hibridization gap \eqref{hgap} for spin-down electrons in a strip geometry of length $L_x=30$. Model parameters: $J = 1.0\Rightarrow t_a^C=0.25$ (in $t$ and $a$ units). }
\label{tabla}
\end{table}

This noticeable and controllable variation in the gap, together with the spatial segregation and spin-splitting already commented, suggests the design of a novel device architecture: an AM FET in the topological phase. The split-drain FET geometry is sketched in Figure \ref{fig_GapFET}(b). By aligning the transport channel (source-drain axis) along the [110] diagonal of the $\dd_{x^2-y^2}$-wave AM (or along the $X$ axis of the  $\dd_{xy}$ configuration), the intrinsic group velocity mismatch steers spin-$\uparrow$ and spin-$\downarrow$ carriers toward opposite lateral boundaries. By designing the contacts to touch only one side of the ribbon, a pure spin current could be turned ON and OFF using the gate. This is the ``Holy Grail'' of spintronics (see e.g.  \cite{fang2025edgetronicstwodimensionalaltermagnets} for other recent ``edgetronics''  proposals in 2D AMs). 

For the transistor to work at temperature $T$, the finite-size gap must be larger than the thermal energy, $\Delta(L_y) > k_B T$, which is about $26$~meV at room temperature. Assuming a typical hopping scale of $t \approx 1$ eV, this should work in principle for
ultra-narrow ribbons at room temperatures.

It is crucial to distinguish this routing architecture from recent proposals regarding the electrical switching of altermagnets \cite{ChenPRL135_2025}. Conventional electrical switching in altermagnetic/antiferromagnetic spintronic devices relies on injecting currents to generate spin-orbit torques that physically rotate the bulk magnetic domains (Néel vector), a dynamically active and potentially energy-intensive process. In stark contrast, the Split-Drain FET proposed here operates with a static magnetic background. The switching is purely electrostatic, it requires no magnetic switching at all. The gate voltage is used to shift the Fermi level,  while the intrinsic $\dd$-wave group velocity naturally steers the spin-up and spin-down electrons to completely different physical drains; That is, it merely selects which spin-polarized edge channel (HOMO or LUMO) is active, while the intrinsic $\dd$-wave group velocity handles the spatial separation of the current. This eliminates the need to overcome magnetic anisotropy barriers, offering a distinct, static pathway for altermagnetic spintronics and providing a highly energy-efficient, purely electrostatic switching mechanism.

Furthermore, while previous studies like \cite{Parshukov_PhysRevB.111.224406} focus on bulk-driven Hall responses, we leverage the intrinsic anisotropy of the edge transport to propose a concrete Split-Drain FET architecture. This device exploits group-velocity steering and spin-dependent deflection, rather than the standard bulk Hall responses studied by Parshukov et al.  \cite{Parshukov_PhysRevB.111.224406},  offering a practical application of altermagnetic edge states that moves beyond the observation of spectral crossings toward active transport control in spintronic devices.  Therefore, our real-space diagnostics and Split-Drain FET proposal provide a completely different perspective on how to exploit these edge states. Also, while Parshukov et al.  provided a thorough symmetry classification of topological crossings and computed standard $k$-space invariants (Chern numbers) alongside open boundary condition edge spectra, our work provides the methodological novelty of deliberately bypassing standard momentum-space or symmetry-based topological invariants. Instead, we introduce information-theoretic real-space markers: fidelity-susceptibility and IPR.

\section{Infinite Strip Geometry}\label{ribbonsec}

First, we studied the band structure in momentum space $(k_x,k_y)$ and then the crystal structure in real space $(x,y)$ of our AM. Now we turn to a mixed momentum-real-space picture $(k_x,y)$ that interpolates between those previous images. We analyze the structure of edge states in the hybrid space $(k_x,y)$ and their localization properties through IPR calculations.

\subsection{Edge and bulk spectra}

To investigate the edge states, we adopt a ribbon geometry of strip width $L_y=L$, infinite in $x\in (-\infty,\infty)$ and finite in $y=0,\dots,L$ (the site indices in $a$ units). Thus, we employ a mixed representation for our wavefunctions $\psi(k_x, y)$.  We perform a partial Fourier transform
in the spin-$s$ Hamiltonian $\mathcal{H}_s=\int_\mathrm{BZ}d\vk \,c^\dag_{\vk} H_s(\vk)c_{\vk}$ by substituting the annihilation (viz. creation) operators
\begin{equation}
 c_{\vk}=\frac{1}{L}\sum_{y=0}^{L} e^{\ic k_y y}c_{k,y},
\end{equation}
to obtain the tight-binding  Hamiltonian  
\begin{equation}
 \mathcal{H}_s=\sum_{k,y} \mathcal{E}_s(k)c_{k,y}^\dag c_{k,y}+\mathcal{T}c_{k,y}^\dag c_{k,y+1}+\mathcal{T}^\dag c_{k,y+1}^\dag c_{k,y},\label{hamTB}
\end{equation}
in the position $y$ and momentum  $k=k_x$ spaces. The $2\times 2$ matrix $ \mathcal{E}_s(k)$  results from eliminating all terms depending on $k_y$ in  $H_s(\vk)=\bm{d}_s(\vk)\cdot \bm{\tau}$. Those $k_y$-terms then contribute to the off-diagonal matrix  $\mathcal{T}$. Doing so, we have
\be
\mathcal{E}_s=\left(\ba{cc}sJ +2t_a\cos(k_x) & t(1+e^{-i k_x}) \\ t(1+e^{i k_x}) & -sJ -2t_a\cos(k_x)  \ea\right),\quad \mathcal{T}= \left(\ba{cc} -t_a & 0\\ t(1+e^{i k_x}) & t_a\ea\right).
\ee

\begin{figure}[h]
	\begin{center}
		\includegraphics[width=0.45\columnwidth]{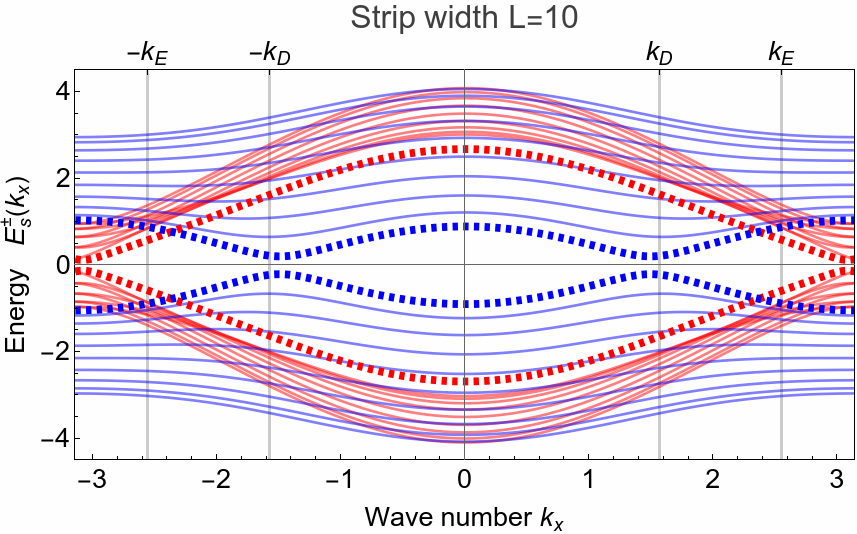}
	\end{center}
	\captionsetup{justification=raggedright, singlelinecheck=false}
	\caption{Energy spectrum $E_s^\pm(k_x)$ for $L=10$ and $t_a>t_a^C$. Bulk (solid) and edge (thick dashed) for spin-up (red) and spin-down (blue) Hamiltonian eigenstates. Model parameters $J=1, t_a=J/2=0.5$ ($t$ and $a$ units). 
	}
	\label{fig_SpectrumL}
\end{figure}

The Hamiltonian matrix $\mathcal{H}_s$ is of size  $2L$ and is numerically diagonalized. Figure \ref{fig_SpectrumL} shows the energy spectrum $E_s(k_x)$ for a strip width of $L=10$ in the topological phase $t_a>t_a^C$,  composed of both: bulk (solid) and edge (thick dashed) spin-up (red) and spin-down (blue) states. We observe some important differences with the infinite sample spectrum of Figure \ref{fig_BandStructure}. Firstly, Dirac $\pm k_D$ and up-down spin edge-state $k_E$ crossing points are slightly shifted with respect to the $L\to\infty$ case. We also observe the opening of the gap for finite $L$ discussed in previous Section \ref{gapsec} for finite size nanoribbons.

\subsection{Edge state structure}

\begin{figure}[h]
	\begin{center}
		\includegraphics[width=\columnwidth]{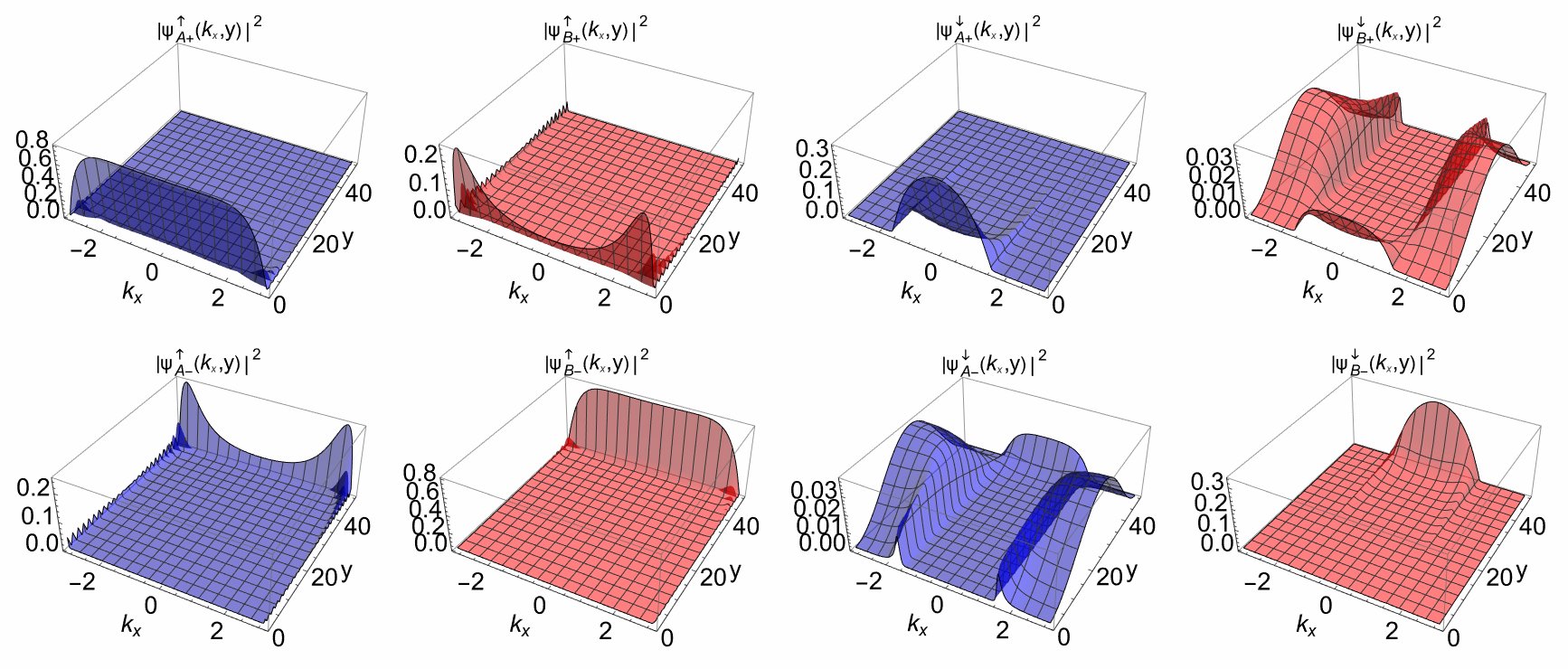}
	\end{center}
	\captionsetup{justification=raggedright, singlelinecheck=false}
	\caption{Four spinor probability density $(|\psi_A^\uparrow|^2,|\psi_B^\uparrow|^2,|\psi_A^\downarrow|^2,|\psi_B^\downarrow|^2)_\pm$ as a function of $(k_x,y)$ for conduction ($+$: top row) and valence ($-$:bottom row) edge staes. Strip width $L=50$. Model parameters:   $J=1, t_a=J/2>t_a^C$ ($t$ and $a$ units).
	}
	\label{fig_DensityBulkEdge2}
\end{figure}

We are interested in the particular structure of edge states in $(k_x,y)$ space. Arranging our  4-spinor states as  $\{\psi_{A}^\uparrow,\psi_B^\uparrow,\psi_A^\downarrow,\psi_B^\downarrow)$, in Figure \ref{fig_DensityBulkEdge2} we plot the probability densities $|\psi_\tau^s(k_x,y)|^2$, for the four components (separately), for conduction (top row) and valence (bottom row) edge states on a strip geometry of width $L=50$. Spin up conduction (valence) electrons are localized at the edge $y=0$ ($y=L$). This is compatible with the ``traffic analogy'' given in Section \ref{LDOSsec}, where we said that spin-up electrons experience a traffic jam in the $Y$-direction, remaining ``parked'' at $y=0,L$. The structure for spin-down electrons is less straightforward to explain. For it, we shall use the IPR introduced in \eqref{IPReq} as a measure of localization.

\subsection{Inverse participation ratio and edge localization}

In our case, the IPR of a spin $s$ 2-spinor $\psi^s=(\psi_{A}^\uparrow,\psi_B^\uparrow)$, normalized as $\int_0^L dy |\psi^s(k_x,y)|^2=1$,  in  position $y$ space for each value of the momentum $k_x$ is given by
\begin{equation}
 \mathrm{IPR}_s(k_x)=\int_0^L dy |\psi^s(k_x,y)|^4, 
\end{equation}
where we understand $|\psi^s|^4=|\psi_A^{s}|^4+|\psi_B^{s}|^4$. 

\begin{figure}[h]
	\begin{center}
		\includegraphics[width=0.7\columnwidth]{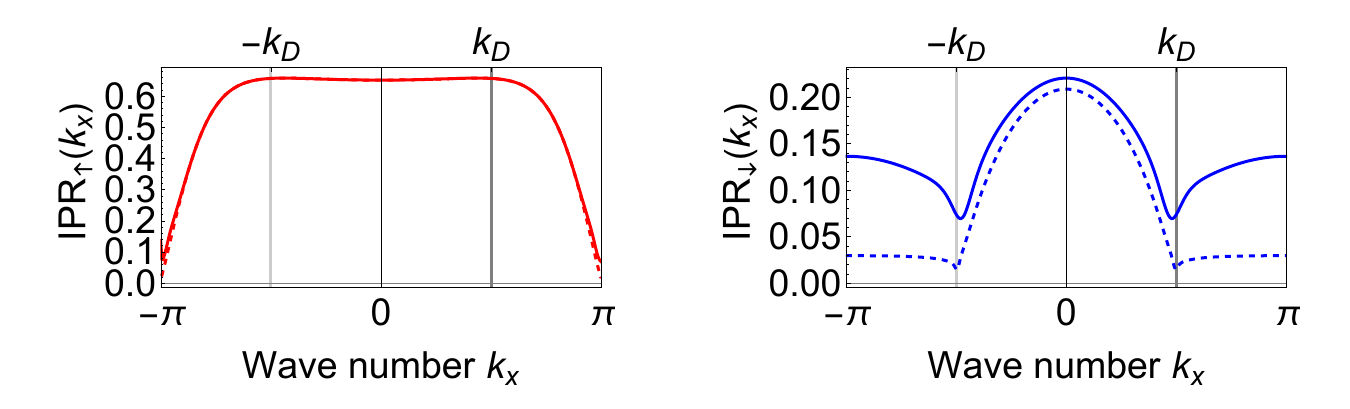}
	\end{center}
	\captionsetup{justification=raggedright, singlelinecheck=false}
	\caption{IPR of spin-up (red) and spin-down (blue) edge conduction states for $L=10$ (solid) and $L=50$ (dashed). Model parameters: $J=1, t_a=J/2>t_a^C$ ($t$ and $a$ units).
	}
	\label{fig_IPRk}
\end{figure}

Figure \ref{fig_IPRk} displays $\mathrm{IPR}_s(k_x)$  (valence and conduction coincide) for spin-up (red) and spin-down (blue) edge states for $L=10$ (solid) and $L=50$ (dashed). Edge states are more localized in space $y\in [0,L]$ for low momentum $|k_x|<k_D$. Spin-down edge states become increasingly delocalized at the Dirac points, especially for larger strip widths $L$, where they delocalize (transmute to bulk states) for $|k_x|>k_D$. Spin-up electrons are highly localized in momentum space, except near $k_x\simeq \pm \pi$, where they delocalize (merge into the bulk).

\section{Comments and Conclusions}\label{conclusec}

The field of altermagnetism has rapidly transitioned from a theoretical curiosity to a cornerstone of modern condensed matter physics. Such a rapid shift may leave aspects of the theoretical foundation insufficiently examined or explained, motivating a deeper analysis. The literature is enormous (see e.g. \cite{JUNGWIRTH2025100162,Fukaya_2025} for recent reviews), but it is sometimes too specialized, focusing more on the experimental aspect than on the theoretical one. This study aims to contribute to ``filling this gap'', by also looking at the problem from a different (information-theoretic) point of view.

We have seen that the AM undergoes a topological semimetal transition at the critical value $t_a^C = J/4$ of the intra-sublattice AM hopping strength $t_a$. This is a transition from a gapped AM insulator to a gapless nodal AM (a semimetal with Dirac points). In the BI phase, the exchange splitting $J$ is strong enough to keep the bands separated everywhere. The AM modulation is too weak to overcome the gap; thus, the system is an insulator (assuming half-filling). At the transition point ($t_{a} = J/4$), the bands touch at a single point $\vk_D=(0,\pi)$ for spin up and $\vk_D=(\pi, 0)$ for spin down. In the TI phase, the modulation is strong enough to invert the bands. The band touching point splits into two Dirac nodes that move along the $k_y = \pi$ ($k_x=\pi$) line for spin down (up) as $t_{a}$ increases. These nodes are topologically protected (they carry a Berry phase of $\pi$).

We also studied the effective spin polarization. It is not just about having a different number of up and down spin states; rather, it is about the motion asymmetry. Although the energy is $\dd$-wave symmetric, the Fermi surface is $\dd$-wave distorted and, consequently, the group velocity is $\dd$-wave steered. For $\dd_{x^2-y^2}$-wave AMs, spin-up (down) electrons preferentially move along the horizontal (vertical) axis. This distinguishes AMs from FMs, where the spin current is usually longitudinal (i.e., parallel to the charge current). Therefore, AM materials act as a ``spin pump'' or as a ``spin-filter''. We also calculated the effective AM strength $P_\mathrm{eff}$ quantifying how much ``usable'' spin is available for transport for a given $t_a$. For our Hamiltonian parameter choice, values of $P_\mathrm{eff}\simeq 80\%$ can be attained in the topological phase.

In a finite size $L_xL_y$ nanoribbon (which is what we care about in technology), spin polarization manifests as spin accumulation at the boundaries. This is the origin of the spin-Hall effect in these materials and the hallmark of the spin-splitter effect (or crystal-Hall effect) in this specific geometry. The  abrupt change in the structure of edge states at the critical point $t_a^C$ is captured by fidelity-susceptibility, which is related to Fisher information and sensing precision though the quantum Cramer-Rao bound. This could have consequences in the design of quantum sensors of topological nature. IPR measures provide a clear distinction between trivial and topological phases. 

Also, by exploiting the hybridization gap $\Delta(L_y)$ for ultrathin ribbons, we have proposed a topological AM field-effect transistor device. Unlike conventional semiconductors,  this device's ON state relies on ballistic edge transport, minimizing scattering. Furthermore, due to the intrinsic spin-momentum locking of the AM edges, such a device would inherently function as a spin-valve, controlling spin-polarized currents via electrostatic gating, provided the ribbon width $L_y$ is sufficiently small (smaller than the coherence length $\lambda$ of edge states) to open a gap exceeding thermal broadening.

Our real-space approach is highly relevant to recent proposals for realizing $\dd$-wave altermagnetism in ultracold Fermi-Hubbard optical lattices \cite{Das_PhysRevLett.132.263402}. In such experiments, quantum gas microscopes provide direct, single-site resolution of the atomic density. Traditional $k$-space topological invariants are difficult to measure in these finite, harmonically trapped systems. However, the IPR and real-space edge localization we calculate here can be directly imaged in cold-atom setups, offering a straightforward experimental pathway to verify the topological phase transition.
    
We also discussed the infinite strip geometry and analyzed the localization of edge states with the inverse participation ratio. This entropic measure indicates how delocalized the spin-up or spin-down electron is in real-space $y$ for a given momentum $k_x$.

Our next step will be the inclusion of spin-orbit couplings and their influence on the topological behavior of AMs, opening up the possibility to other g- i- or p-wave symmetries. Also adding a stacking degree of freedom by going from one layer to multilayer arrangements, which increases the range of new technological possibilities regarding monolayer quantum devices. Such is the case of the already mentioned exotic  physics of twisted bilayer graphene \cite{Cao2018,Cao2018unconventional,TBGPhysicaE}. Also recent studies on optically tunable spin transport in bilayer altermagnetic Mott insulators \cite{sicheler2025opticallytunablespintransport}. This is left for future work.

\begin{acknowledgments}
I thank the support of the Spanish Ministry of Science, Innovation and Universities through the
project PID2022-138144NB-I00. I also acknowledge support from PAIDI group FQM-420 of the University of Granada.
\end{acknowledgments}

\bibliography{bibliografia.bib}

\begin{thebibliography}{42}%
\makeatletter
\providecommand \@ifxundefined [1]{%
 \@ifx{#1\undefined}
}%
\providecommand \@ifnum [1]{%
 \ifnum #1\expandafter \@firstoftwo
 \else \expandafter \@secondoftwo
 \fi
}%
\providecommand \@ifx [1]{%
 \ifx #1\expandafter \@firstoftwo
 \else \expandafter \@secondoftwo
 \fi
}%
\providecommand \natexlab [1]{#1}%
\providecommand \enquote  [1]{``#1''}%
\providecommand \bibnamefont  [1]{#1}%
\providecommand \bibfnamefont [1]{#1}%
\providecommand \citenamefont [1]{#1}%
\providecommand \href@noop [0]{\@secondoftwo}%
\providecommand \href [0]{\begingroup \@sanitize@url \@href}%
\providecommand \@href[1]{\@@startlink{#1}\@@href}%
\providecommand \@@href[1]{\endgroup#1\@@endlink}%
\providecommand \@sanitize@url [0]{\catcode `\\12\catcode `\$12\catcode
  `\&12\catcode `\#12\catcode `\^12\catcode `\_12\catcode `\%12\relax}%
\providecommand \@@startlink[1]{}%
\providecommand \@@endlink[0]{}%
\providecommand \url  [0]{\begingroup\@sanitize@url \@url }%
\providecommand \@url [1]{\endgroup\@href {#1}{\urlprefix }}%
\providecommand \urlprefix  [0]{URL }%
\providecommand \Eprint [0]{\href }%
\providecommand \doibase [0]{https://doi.org/}%
\providecommand \selectlanguage [0]{\@gobble}%
\providecommand \bibinfo  [0]{\@secondoftwo}%
\providecommand \bibfield  [0]{\@secondoftwo}%
\providecommand \translation [1]{[#1]}%
\providecommand \BibitemOpen [0]{}%
\providecommand \bibitemStop [0]{}%
\providecommand \bibitemNoStop [0]{.\EOS\space}%
\providecommand \EOS [0]{\spacefactor3000\relax}%
\providecommand \BibitemShut  [1]{\csname bibitem#1\endcsname}%
\let\auto@bib@innerbib\@empty
\bibitem [{\citenamefont {Mazin}(2022)}]{Mazin2022_Editorial}%
  \BibitemOpen
  \bibfield  {author} {\bibinfo {author} {\bibfnamefont {I.~I.}\ \bibnamefont
  {Mazin}},\ }\bibfield  {title} {\bibinfo {title} {Editorial:
  Altermagnetism—a new punch line of fundamental magnetism},\ }\href
  {https://doi.org/10.1103/PhysRevX.12.040002} {\bibfield  {journal} {\bibinfo
  {journal} {Phys. Rev. X}\ }\textbf {\bibinfo {volume} {12}},\ \bibinfo
  {pages} {040002} (\bibinfo {year} {2022})}\BibitemShut {NoStop}%
\bibitem [{\citenamefont {\v{S}mejkal}\ \emph
  {et~al.}(2022{\natexlab{a}})\citenamefont {\v{S}mejkal}, \citenamefont
  {Sinova},\ and\ \citenamefont {Jungwirth}}]{Smejkal2022_Theory}%
  \BibitemOpen
  \bibfield  {author} {\bibinfo {author} {\bibfnamefont {L.}~\bibnamefont
  {\v{S}mejkal}}, \bibinfo {author} {\bibfnamefont {J.}~\bibnamefont
  {Sinova}},\ and\ \bibinfo {author} {\bibfnamefont {T.}~\bibnamefont
  {Jungwirth}},\ }\bibfield  {title} {\bibinfo {title} {Beyond conventional
  ferromagnetism and antiferromagnetism: A phase with nonrelativistic spin and
  crystal rotation symmetry},\ }\href
  {https://doi.org/10.1103/PhysRevX.12.031042} {\bibfield  {journal} {\bibinfo
  {journal} {Phys. Rev. X}\ }\textbf {\bibinfo {volume} {12}},\ \bibinfo
  {pages} {031042} (\bibinfo {year} {2022}{\natexlab{a}})}\BibitemShut
  {NoStop}%
\bibitem [{\citenamefont {\v{S}mejkal}\ \emph
  {et~al.}(2022{\natexlab{b}})\citenamefont {\v{S}mejkal}, \citenamefont
  {Sinova},\ and\ \citenamefont {Jungwirth}}]{Smejkal2022_Landscape}%
  \BibitemOpen
  \bibfield  {author} {\bibinfo {author} {\bibfnamefont {L.}~\bibnamefont
  {\v{S}mejkal}}, \bibinfo {author} {\bibfnamefont {J.}~\bibnamefont
  {Sinova}},\ and\ \bibinfo {author} {\bibfnamefont {T.}~\bibnamefont
  {Jungwirth}},\ }\bibfield  {title} {\bibinfo {title} {Emerging research
  landscape of altermagnetism},\ }\href
  {https://doi.org/10.1103/PhysRevX.12.040501} {\bibfield  {journal} {\bibinfo
  {journal} {Phys. Rev. X}\ }\textbf {\bibinfo {volume} {12}},\ \bibinfo
  {pages} {040501} (\bibinfo {year} {2022}{\natexlab{b}})}\BibitemShut
  {NoStop}%
\bibitem [{\citenamefont {Krempask\'y}\ \emph {et~al.}(2024)\citenamefont
  {Krempask\'y}, \citenamefont {\v{S}mejkal}, \citenamefont {D'Souza},
  \citenamefont {Hajlaoui}, \citenamefont {Springholz} \emph
  {et~al.}}]{Krempasky2024_MnTe}%
  \BibitemOpen
  \bibfield  {author} {\bibinfo {author} {\bibfnamefont {J.}~\bibnamefont
  {Krempask\'y}}, \bibinfo {author} {\bibfnamefont {L.}~\bibnamefont
  {\v{S}mejkal}}, \bibinfo {author} {\bibfnamefont {S.~W.}\ \bibnamefont
  {D'Souza}}, \bibinfo {author} {\bibfnamefont {M.}~\bibnamefont {Hajlaoui}},
  \bibinfo {author} {\bibfnamefont {G.}~\bibnamefont {Springholz}}, \emph
  {et~al.},\ }\bibfield  {title} {\bibinfo {title} {Altermagnetic lifting of
  {Kramers} spin degeneracy},\ }\href
  {https://doi.org/10.1038/s41586-023-06907-7} {\bibfield  {journal} {\bibinfo
  {journal} {Nature}\ }\textbf {\bibinfo {volume} {626}},\ \bibinfo {pages}
  {517} (\bibinfo {year} {2024})}\BibitemShut {NoStop}%
\bibitem [{\citenamefont {Schiff}\ \emph {et~al.}(2025)\citenamefont {Schiff},
  \citenamefont {Corticelli}, \citenamefont {Guerreiro}, \citenamefont
  {Romhányi},\ and\ \citenamefont {McClarty}}]{10.21468/SciPostPhys.18.3.109}%
  \BibitemOpen
  \bibfield  {author} {\bibinfo {author} {\bibfnamefont {H.}~\bibnamefont
  {Schiff}}, \bibinfo {author} {\bibfnamefont {A.}~\bibnamefont {Corticelli}},
  \bibinfo {author} {\bibfnamefont {A.}~\bibnamefont {Guerreiro}}, \bibinfo
  {author} {\bibfnamefont {J.}~\bibnamefont {Romhányi}},\ and\ \bibinfo
  {author} {\bibfnamefont {P.}~\bibnamefont {McClarty}},\ }\bibfield  {title}
  {\bibinfo {title} {{The crystallographic spin point groups and their
  representations}},\ }\href {https://doi.org/10.21468/SciPostPhys.18.3.109}
  {\bibfield  {journal} {\bibinfo  {journal} {SciPost Phys.}\ }\textbf
  {\bibinfo {volume} {18}},\ \bibinfo {pages} {109} (\bibinfo {year}
  {2025})}\BibitemShut {NoStop}%
\bibitem [{\citenamefont {Roig}\ \emph {et~al.}(2024)\citenamefont {Roig},
  \citenamefont {Kreisel}, \citenamefont {Yu}, \citenamefont {Andersen},\ and\
  \citenamefont {Agterberg}}]{Roig_PhysRevB.110.144412}%
  \BibitemOpen
  \bibfield  {author} {\bibinfo {author} {\bibfnamefont {M.}~\bibnamefont
  {Roig}}, \bibinfo {author} {\bibfnamefont {A.}~\bibnamefont {Kreisel}},
  \bibinfo {author} {\bibfnamefont {Y.}~\bibnamefont {Yu}}, \bibinfo {author}
  {\bibfnamefont {B.~M.}\ \bibnamefont {Andersen}},\ and\ \bibinfo {author}
  {\bibfnamefont {D.~F.}\ \bibnamefont {Agterberg}},\ }\bibfield  {title}
  {\bibinfo {title} {Minimal models for altermagnetism},\ }\href
  {https://doi.org/10.1103/PhysRevB.110.144412} {\bibfield  {journal} {\bibinfo
   {journal} {Phys. Rev. B}\ }\textbf {\bibinfo {volume} {110}},\ \bibinfo
  {pages} {144412} (\bibinfo {year} {2024})}\BibitemShut {NoStop}%
\bibitem [{\citenamefont {Mazin}(2023)}]{Mazin2023_Notes}%
  \BibitemOpen
  \bibfield  {author} {\bibinfo {author} {\bibfnamefont {I.~I.}\ \bibnamefont
  {Mazin}},\ }\bibfield  {title} {\bibinfo {title} {Altermagnetism in {MnTe}:
  Origin, predicted manifestations, and routes to detwinning},\ }\href
  {https://doi.org/10.1103/PhysRevB.107.L100418} {\bibfield  {journal}
  {\bibinfo  {journal} {Phys. Rev. B}\ }\textbf {\bibinfo {volume} {107}},\
  \bibinfo {pages} {L100418} (\bibinfo {year} {2023})}\BibitemShut {NoStop}%
\bibitem [{\citenamefont {Feng}\ \emph {et~al.}(2022)\citenamefont {Feng},
  \citenamefont {Zhou}, \citenamefont {\v{S}mejkal}, \citenamefont {Wu},
  \citenamefont {Zhu}, \citenamefont {Guo},\ and\ \citenamefont
  {Yao}}]{Feng2022_RuO2_AHE}%
  \BibitemOpen
  \bibfield  {author} {\bibinfo {author} {\bibfnamefont {W.}~\bibnamefont
  {Feng}}, \bibinfo {author} {\bibfnamefont {X.}~\bibnamefont {Zhou}}, \bibinfo
  {author} {\bibfnamefont {L.}~\bibnamefont {\v{S}mejkal}}, \bibinfo {author}
  {\bibfnamefont {L.}~\bibnamefont {Wu}}, \bibinfo {author} {\bibfnamefont
  {Z.}~\bibnamefont {Zhu}}, \bibinfo {author} {\bibfnamefont {H.}~\bibnamefont
  {Guo}},\ and\ \bibinfo {author} {\bibfnamefont {Y.}~\bibnamefont {Yao}},\
  }\bibfield  {title} {\bibinfo {title} {An anomalous hall effect in
  altermagnetic ruthenium dioxide},\ }\href
  {https://doi.org/10.1038/s41928-022-00866-z} {\bibfield  {journal} {\bibinfo
  {journal} {Nat. Electron.}\ }\textbf {\bibinfo {volume} {5}},\ \bibinfo
  {pages} {735} (\bibinfo {year} {2022})}\BibitemShut {NoStop}%
\bibitem [{\citenamefont {Fedchenko}\ \emph {et~al.}(2024)\citenamefont
  {Fedchenko}, \citenamefont {Min\'ar}, \citenamefont {Akashdeep},
  \citenamefont {D'Souza} \emph {et~al.}}]{Fedchenko2024_RuO2_ARPES}%
  \BibitemOpen
  \bibfield  {author} {\bibinfo {author} {\bibfnamefont {O.}~\bibnamefont
  {Fedchenko}}, \bibinfo {author} {\bibfnamefont {J.}~\bibnamefont {Min\'ar}},
  \bibinfo {author} {\bibfnamefont {A.}~\bibnamefont {Akashdeep}}, \bibinfo
  {author} {\bibfnamefont {S.~W.}\ \bibnamefont {D'Souza}}, \emph {et~al.},\
  }\bibfield  {title} {\bibinfo {title} {Observation of time-reversal symmetry
  breaking in the band structure of altermagnetic {RuO$_2$}},\ }\href
  {https://doi.org/10.1126/sciadv.adj4883} {\bibfield  {journal} {\bibinfo
  {journal} {Sci. Adv.}\ }\textbf {\bibinfo {volume} {10}},\ \bibinfo {pages}
  {eadj4883} (\bibinfo {year} {2024})}\BibitemShut {NoStop}%
\bibitem [{\citenamefont {Liu}\ \emph {et~al.}(2024)\citenamefont {Liu},
  \citenamefont {Zhan}, \citenamefont {Li}, \citenamefont {Liu}, \citenamefont
  {Cheng}, \citenamefont {Shi}, \citenamefont {Deng}, \citenamefont {Zhang},
  \citenamefont {Li}, \citenamefont {Ding}, \citenamefont {Jiang},
  \citenamefont {Ye}, \citenamefont {Liu}, \citenamefont {Jiang}, \citenamefont
  {Wang}, \citenamefont {Li}, \citenamefont {Xie}, \citenamefont {Wang},
  \citenamefont {Qiao}, \citenamefont {Wen}, \citenamefont {Sun},\ and\
  \citenamefont {Shen}}]{PhysRevLett.133.176401}%
  \BibitemOpen
  \bibfield  {author} {\bibinfo {author} {\bibfnamefont {J.}~\bibnamefont
  {Liu}}, \bibinfo {author} {\bibfnamefont {J.}~\bibnamefont {Zhan}}, \bibinfo
  {author} {\bibfnamefont {T.}~\bibnamefont {Li}}, \bibinfo {author}
  {\bibfnamefont {J.}~\bibnamefont {Liu}}, \bibinfo {author} {\bibfnamefont
  {S.}~\bibnamefont {Cheng}}, \bibinfo {author} {\bibfnamefont
  {Y.}~\bibnamefont {Shi}}, \bibinfo {author} {\bibfnamefont {L.}~\bibnamefont
  {Deng}}, \bibinfo {author} {\bibfnamefont {M.}~\bibnamefont {Zhang}},
  \bibinfo {author} {\bibfnamefont {C.}~\bibnamefont {Li}}, \bibinfo {author}
  {\bibfnamefont {J.}~\bibnamefont {Ding}}, \bibinfo {author} {\bibfnamefont
  {Q.}~\bibnamefont {Jiang}}, \bibinfo {author} {\bibfnamefont
  {M.}~\bibnamefont {Ye}}, \bibinfo {author} {\bibfnamefont {Z.}~\bibnamefont
  {Liu}}, \bibinfo {author} {\bibfnamefont {Z.}~\bibnamefont {Jiang}}, \bibinfo
  {author} {\bibfnamefont {S.}~\bibnamefont {Wang}}, \bibinfo {author}
  {\bibfnamefont {Q.}~\bibnamefont {Li}}, \bibinfo {author} {\bibfnamefont
  {Y.}~\bibnamefont {Xie}}, \bibinfo {author} {\bibfnamefont {Y.}~\bibnamefont
  {Wang}}, \bibinfo {author} {\bibfnamefont {S.}~\bibnamefont {Qiao}}, \bibinfo
  {author} {\bibfnamefont {J.}~\bibnamefont {Wen}}, \bibinfo {author}
  {\bibfnamefont {Y.}~\bibnamefont {Sun}},\ and\ \bibinfo {author}
  {\bibfnamefont {D.}~\bibnamefont {Shen}},\ }\bibfield  {title} {\bibinfo
  {title} {Absence of altermagnetic spin splitting character in rutile oxide
  ${\mathrm{ruo}}_{2}$},\ }\href
  {https://doi.org/10.1103/PhysRevLett.133.176401} {\bibfield  {journal}
  {\bibinfo  {journal} {Phys. Rev. Lett.}\ }\textbf {\bibinfo {volume} {133}},\
  \bibinfo {pages} {176401} (\bibinfo {year} {2024})}\BibitemShut {NoStop}%
\bibitem [{\citenamefont {Morano}\ \emph {et~al.}(2025)\citenamefont {Morano},
  \citenamefont {Maesen}, \citenamefont {Nikitin}, \citenamefont {Lass},
  \citenamefont {Mazzone},\ and\ \citenamefont
  {Zaharko}}]{PhysRevLett.134.226702}%
  \BibitemOpen
  \bibfield  {author} {\bibinfo {author} {\bibfnamefont {V.~C.}\ \bibnamefont
  {Morano}}, \bibinfo {author} {\bibfnamefont {Z.}~\bibnamefont {Maesen}},
  \bibinfo {author} {\bibfnamefont {S.~E.}\ \bibnamefont {Nikitin}}, \bibinfo
  {author} {\bibfnamefont {J.}~\bibnamefont {Lass}}, \bibinfo {author}
  {\bibfnamefont {D.~G.}\ \bibnamefont {Mazzone}},\ and\ \bibinfo {author}
  {\bibfnamefont {O.}~\bibnamefont {Zaharko}},\ }\bibfield  {title} {\bibinfo
  {title} {Absence of altermagnetic magnon band splitting in
  ${\mathrm{mnf}}_{2}$},\ }\href
  {https://doi.org/10.1103/PhysRevLett.134.226702} {\bibfield  {journal}
  {\bibinfo  {journal} {Phys. Rev. Lett.}\ }\textbf {\bibinfo {volume} {134}},\
  \bibinfo {pages} {226702} (\bibinfo {year} {2025})}\BibitemShut {NoStop}%
\bibitem [{\citenamefont {Zeng}\ \emph {et~al.}(2025)\citenamefont {Zeng},
  \citenamefont {Liu}, \citenamefont {Zhu}, \citenamefont {Zheng},
  \citenamefont {Liu}, \citenamefont {Zhu}, \citenamefont {Shao}, \citenamefont
  {Hao}, \citenamefont {Ma}, \citenamefont {Qu}, \citenamefont {Kurleto},
  \citenamefont {Wutke}, \citenamefont {Luo}, \citenamefont {Dai},
  \citenamefont {Zhang}, \citenamefont {Miyamoto}, \citenamefont {Shimada},
  \citenamefont {Okuda}, \citenamefont {Tanaka}, \citenamefont {Huang},
  \citenamefont {Liu},\ and\ \citenamefont
  {Liu}}]{zeng2025nonaltermagneticspintexturemnte}%
  \BibitemOpen
  \bibfield  {author} {\bibinfo {author} {\bibfnamefont {M.}~\bibnamefont
  {Zeng}}, \bibinfo {author} {\bibfnamefont {P.}~\bibnamefont {Liu}}, \bibinfo
  {author} {\bibfnamefont {M.-Y.}\ \bibnamefont {Zhu}}, \bibinfo {author}
  {\bibfnamefont {N.}~\bibnamefont {Zheng}}, \bibinfo {author} {\bibfnamefont
  {X.-R.}\ \bibnamefont {Liu}}, \bibinfo {author} {\bibfnamefont {Y.-P.}\
  \bibnamefont {Zhu}}, \bibinfo {author} {\bibfnamefont {T.-H.}\ \bibnamefont
  {Shao}}, \bibinfo {author} {\bibfnamefont {Y.-J.}\ \bibnamefont {Hao}},
  \bibinfo {author} {\bibfnamefont {X.-M.}\ \bibnamefont {Ma}}, \bibinfo
  {author} {\bibfnamefont {G.}~\bibnamefont {Qu}}, \bibinfo {author}
  {\bibfnamefont {R.}~\bibnamefont {Kurleto}}, \bibinfo {author} {\bibfnamefont
  {D.}~\bibnamefont {Wutke}}, \bibinfo {author} {\bibfnamefont {R.-H.}\
  \bibnamefont {Luo}}, \bibinfo {author} {\bibfnamefont {Y.}~\bibnamefont
  {Dai}}, \bibinfo {author} {\bibfnamefont {X.}~\bibnamefont {Zhang}}, \bibinfo
  {author} {\bibfnamefont {K.}~\bibnamefont {Miyamoto}}, \bibinfo {author}
  {\bibfnamefont {K.}~\bibnamefont {Shimada}}, \bibinfo {author} {\bibfnamefont
  {T.}~\bibnamefont {Okuda}}, \bibinfo {author} {\bibfnamefont
  {K.}~\bibnamefont {Tanaka}}, \bibinfo {author} {\bibfnamefont
  {Y.}~\bibnamefont {Huang}}, \bibinfo {author} {\bibfnamefont
  {Q.}~\bibnamefont {Liu}},\ and\ \bibinfo {author} {\bibfnamefont
  {C.}~\bibnamefont {Liu}},\ }\href {https://arxiv.org/abs/2511.02447}
  {\bibinfo {title} {Non-altermagnetic spin texture in {MnTe}}} (\bibinfo
  {year} {2025}),\ \Eprint {https://arxiv.org/abs/2511.02447} {arXiv:2511.02447
  [cond-mat.mtrl-sci]} \BibitemShut {NoStop}%
\bibitem [{\citenamefont {Das}\ \emph {et~al.}(2024)\citenamefont {Das},
  \citenamefont {Leeb}, \citenamefont {Knolle},\ and\ \citenamefont
  {Knap}}]{Das_PhysRevLett.132.263402}%
  \BibitemOpen
  \bibfield  {author} {\bibinfo {author} {\bibfnamefont {P.}~\bibnamefont
  {Das}}, \bibinfo {author} {\bibfnamefont {V.}~\bibnamefont {Leeb}}, \bibinfo
  {author} {\bibfnamefont {J.}~\bibnamefont {Knolle}},\ and\ \bibinfo {author}
  {\bibfnamefont {M.}~\bibnamefont {Knap}},\ }\bibfield  {title} {\bibinfo
  {title} {Realizing altermagnetism in fermi-hubbard models with ultracold
  atoms},\ }\href {https://doi.org/10.1103/PhysRevLett.132.263402} {\bibfield
  {journal} {\bibinfo  {journal} {Phys. Rev. Lett.}\ }\textbf {\bibinfo
  {volume} {132}},\ \bibinfo {pages} {263402} (\bibinfo {year}
  {2024})}\BibitemShut {NoStop}%
\bibitem [{\citenamefont {Manchon}\ and\ \citenamefont
  {Zhang}(2008)}]{PhysRevB.78.212405}%
  \BibitemOpen
  \bibfield  {author} {\bibinfo {author} {\bibfnamefont {A.}~\bibnamefont
  {Manchon}}\ and\ \bibinfo {author} {\bibfnamefont {S.}~\bibnamefont
  {Zhang}},\ }\bibfield  {title} {\bibinfo {title} {Theory of nonequilibrium
  intrinsic spin torque in a single nanomagnet},\ }\href
  {https://doi.org/10.1103/PhysRevB.78.212405} {\bibfield  {journal} {\bibinfo
  {journal} {Phys. Rev. B}\ }\textbf {\bibinfo {volume} {78}},\ \bibinfo
  {pages} {212405} (\bibinfo {year} {2008})}\BibitemShut {NoStop}%
\bibitem [{\citenamefont {Miron}\ \emph {et~al.}(2011)\citenamefont {Miron},
  \citenamefont {Garello}, \citenamefont {Gaudin}, \citenamefont {Zermatten},
  \citenamefont {Costache}, \citenamefont {Auffret}, \citenamefont {Bandiera},
  \citenamefont {Rodmacq}, \citenamefont {Schuhl},\ and\ \citenamefont
  {Gambardella}}]{Miron2011}%
  \BibitemOpen
  \bibfield  {author} {\bibinfo {author} {\bibfnamefont {I.~M.}\ \bibnamefont
  {Miron}}, \bibinfo {author} {\bibfnamefont {K.}~\bibnamefont {Garello}},
  \bibinfo {author} {\bibfnamefont {G.}~\bibnamefont {Gaudin}}, \bibinfo
  {author} {\bibfnamefont {P.-J.}\ \bibnamefont {Zermatten}}, \bibinfo {author}
  {\bibfnamefont {M.~V.}\ \bibnamefont {Costache}}, \bibinfo {author}
  {\bibfnamefont {S.}~\bibnamefont {Auffret}}, \bibinfo {author} {\bibfnamefont
  {S.}~\bibnamefont {Bandiera}}, \bibinfo {author} {\bibfnamefont
  {B.}~\bibnamefont {Rodmacq}}, \bibinfo {author} {\bibfnamefont
  {A.}~\bibnamefont {Schuhl}},\ and\ \bibinfo {author} {\bibfnamefont
  {P.}~\bibnamefont {Gambardella}},\ }\bibfield  {title} {\bibinfo {title}
  {Perpendicular switching of a single ferromagnetic layer induced by in-plane
  current injection},\ }\href {https://doi.org/10.1038/nature10309} {\bibfield
  {journal} {\bibinfo  {journal} {Nature}\ }\textbf {\bibinfo {volume} {476}},\
  \bibinfo {pages} {189} (\bibinfo {year} {2011})}\BibitemShut {NoStop}%
\bibitem [{\citenamefont {Liu}\ \emph {et~al.}(2012)\citenamefont {Liu},
  \citenamefont {Pai}, \citenamefont {Li}, \citenamefont {Tseng}, \citenamefont
  {Ralph},\ and\ \citenamefont {Buhrman}}]{doi:10.1126/science.1218197}%
  \BibitemOpen
  \bibfield  {author} {\bibinfo {author} {\bibfnamefont {L.}~\bibnamefont
  {Liu}}, \bibinfo {author} {\bibfnamefont {C.-F.}\ \bibnamefont {Pai}},
  \bibinfo {author} {\bibfnamefont {Y.}~\bibnamefont {Li}}, \bibinfo {author}
  {\bibfnamefont {H.~W.}\ \bibnamefont {Tseng}}, \bibinfo {author}
  {\bibfnamefont {D.~C.}\ \bibnamefont {Ralph}},\ and\ \bibinfo {author}
  {\bibfnamefont {R.~A.}\ \bibnamefont {Buhrman}},\ }\bibfield  {title}
  {\bibinfo {title} {Spin-torque switching with the giant spin {Hall} effect of
  tantalum},\ }\href {https://doi.org/10.1126/science.1218197} {\bibfield
  {journal} {\bibinfo  {journal} {Science}\ }\textbf {\bibinfo {volume}
  {336}},\ \bibinfo {pages} {555–558} (\bibinfo {year} {2012})}\BibitemShut
  {NoStop}%
\bibitem [{\citenamefont {Gonz\'alez-Hern\'andez}\ \emph
  {et~al.}(2021)\citenamefont {Gonz\'alez-Hern\'andez}, \citenamefont
  {\ifmmode~\check{S}\else \v{S}\fi{}mejkal}, \citenamefont {V\'yborn\'y},
  \citenamefont {Yahagi}, \citenamefont {Sinova}, \citenamefont {Jungwirth},\
  and\ \citenamefont {\ifmmode~\check{Z}\else
  \v{Z}\fi{}elezn\'y}}]{PhysRevLett.126.127701}%
  \BibitemOpen
  \bibfield  {author} {\bibinfo {author} {\bibfnamefont {R.}~\bibnamefont
  {Gonz\'alez-Hern\'andez}}, \bibinfo {author} {\bibfnamefont {L.}~\bibnamefont
  {\ifmmode~\check{S}\else \v{S}\fi{}mejkal}}, \bibinfo {author} {\bibfnamefont
  {K.}~\bibnamefont {V\'yborn\'y}}, \bibinfo {author} {\bibfnamefont
  {Y.}~\bibnamefont {Yahagi}}, \bibinfo {author} {\bibfnamefont
  {J.}~\bibnamefont {Sinova}}, \bibinfo {author} {\bibfnamefont {T.~c.~v.}\
  \bibnamefont {Jungwirth}},\ and\ \bibinfo {author} {\bibfnamefont
  {J.}~\bibnamefont {\ifmmode~\check{Z}\else \v{Z}\fi{}elezn\'y}},\ }\bibfield
  {title} {\bibinfo {title} {Efficient electrical spin splitter based on
  nonrelativistic collinear antiferromagnetism},\ }\href
  {https://doi.org/10.1103/PhysRevLett.126.127701} {\bibfield  {journal}
  {\bibinfo  {journal} {Phys. Rev. Lett.}\ }\textbf {\bibinfo {volume} {126}},\
  \bibinfo {pages} {127701} (\bibinfo {year} {2021})}\BibitemShut {NoStop}%
\bibitem [{\citenamefont {Bai}\ \emph {et~al.}(2021)\citenamefont {Bai},
  \citenamefont {Han}, \citenamefont {Feng}, \citenamefont {Zhou},
  \citenamefont {Su}, \citenamefont {Wang}, \citenamefont {Liao}, \citenamefont
  {Zhu}, \citenamefont {Chen}, \citenamefont {Pan}, \citenamefont {Fan},\ and\
  \citenamefont {Song}}]{Bai2021ObservationOS}%
  \BibitemOpen
  \bibfield  {author} {\bibinfo {author} {\bibfnamefont {H.}~\bibnamefont
  {Bai}}, \bibinfo {author} {\bibfnamefont {L.}~\bibnamefont {Han}}, \bibinfo
  {author} {\bibfnamefont {X.}~\bibnamefont {Feng}}, \bibinfo {author}
  {\bibfnamefont {Y.~J.}\ \bibnamefont {Zhou}}, \bibinfo {author}
  {\bibfnamefont {R.}~\bibnamefont {Su}}, \bibinfo {author} {\bibfnamefont
  {Q.}~\bibnamefont {Wang}}, \bibinfo {author} {\bibfnamefont {L.}~\bibnamefont
  {Liao}}, \bibinfo {author} {\bibfnamefont {W.}~\bibnamefont {Zhu}}, \bibinfo
  {author} {\bibfnamefont {X.}~\bibnamefont {Chen}}, \bibinfo {author}
  {\bibfnamefont {F.}~\bibnamefont {Pan}}, \bibinfo {author} {\bibfnamefont
  {X.}~\bibnamefont {Fan}},\ and\ \bibinfo {author} {\bibfnamefont
  {C.}~\bibnamefont {Song}},\ }\bibfield  {title} {\bibinfo {title}
  {Observation of spin splitting torque in a collinear antiferromagnet
  {RuO${}_2$}.},\ }\href {https://api.semanticscholar.org/CorpusID:237493733}
  {\bibfield  {journal} {\bibinfo  {journal} {Phys. Rev. Lett.}\ }\textbf
  {\bibinfo {volume} {128 19}},\ \bibinfo {pages} {197202} (\bibinfo {year}
  {2021})}\BibitemShut {NoStop}%
\bibitem [{\citenamefont {Nguyen}\ \emph {et~al.}(2024)\citenamefont {Nguyen},
  \citenamefont {Rao}, \citenamefont {Wostyn},\ and\ \citenamefont
  {Couet}}]{Nguyen2024}%
  \BibitemOpen
  \bibfield  {author} {\bibinfo {author} {\bibfnamefont {V.~D.}\ \bibnamefont
  {Nguyen}}, \bibinfo {author} {\bibfnamefont {S.}~\bibnamefont {Rao}},
  \bibinfo {author} {\bibfnamefont {K.}~\bibnamefont {Wostyn}},\ and\ \bibinfo
  {author} {\bibfnamefont {S.}~\bibnamefont {Couet}},\ }\bibfield  {title}
  {\bibinfo {title} {Recent progress in spin-orbit torque magnetic
  random-access memory},\ }\href {https://doi.org/10.1038/s44306-024-00044-1}
  {\bibfield  {journal} {\bibinfo  {journal} {npj Spintronics}\ }\textbf
  {\bibinfo {volume} {2}},\ \bibinfo {pages} {48} (\bibinfo {year}
  {2024})}\BibitemShut {NoStop}%
\bibitem [{\citenamefont {Fukaya}\ \emph {et~al.}(2025)\citenamefont {Fukaya},
  \citenamefont {Lu}, \citenamefont {Yada}, \citenamefont {Tanaka},\ and\
  \citenamefont {Cayao}}]{Fukaya_2025}%
  \BibitemOpen
  \bibfield  {author} {\bibinfo {author} {\bibfnamefont {Y.}~\bibnamefont
  {Fukaya}}, \bibinfo {author} {\bibfnamefont {B.}~\bibnamefont {Lu}}, \bibinfo
  {author} {\bibfnamefont {K.}~\bibnamefont {Yada}}, \bibinfo {author}
  {\bibfnamefont {Y.}~\bibnamefont {Tanaka}},\ and\ \bibinfo {author}
  {\bibfnamefont {J.}~\bibnamefont {Cayao}},\ }\bibfield  {title} {\bibinfo
  {title} {Superconducting phenomena in systems with unconventional magnets},\
  }\href {https://doi.org/10.1088/1361-648X/adf1cf} {\bibfield  {journal}
  {\bibinfo  {journal} {Journal of Physics: Condensed Matter}\ }\textbf
  {\bibinfo {volume} {37}},\ \bibinfo {pages} {313003} (\bibinfo {year}
  {2025})}\BibitemShut {NoStop}%
\bibitem [{\citenamefont {Del~Re}(2025)}]{DelRe7PRR7}%
  \BibitemOpen
  \bibfield  {author} {\bibinfo {author} {\bibfnamefont {L.}~\bibnamefont
  {Del~Re}},\ }\bibfield  {title} {\bibinfo {title} {Dirac points and
  topological phases in correlated altermagnets},\ }\href
  {https://doi.org/10.1103/7nvm-s225} {\bibfield  {journal} {\bibinfo
  {journal} {Phys. Rev. Res.}\ }\textbf {\bibinfo {volume} {7}},\ \bibinfo
  {pages} {033234} (\bibinfo {year} {2025})}\BibitemShut {NoStop}%
\bibitem [{\citenamefont {Attias}\ \emph {et~al.}(2024)\citenamefont {Attias},
  \citenamefont {Levchenko},\ and\ \citenamefont
  {Khodas}}]{PhysRevB.110.094425}%
  \BibitemOpen
  \bibfield  {author} {\bibinfo {author} {\bibfnamefont {L.}~\bibnamefont
  {Attias}}, \bibinfo {author} {\bibfnamefont {A.}~\bibnamefont {Levchenko}},\
  and\ \bibinfo {author} {\bibfnamefont {M.}~\bibnamefont {Khodas}},\
  }\bibfield  {title} {\bibinfo {title} {Intrinsic anomalous hall effect in
  altermagnets},\ }\href {https://doi.org/10.1103/PhysRevB.110.094425}
  {\bibfield  {journal} {\bibinfo  {journal} {Phys. Rev. B}\ }\textbf {\bibinfo
  {volume} {110}},\ \bibinfo {pages} {094425} (\bibinfo {year}
  {2024})}\BibitemShut {NoStop}%
\bibitem [{\citenamefont {Calixto}\ and\ \citenamefont
  {Romera}(2015{\natexlab{a}})}]{silicene1}%
  \BibitemOpen
  \bibfield  {author} {\bibinfo {author} {\bibfnamefont {M.}~\bibnamefont
  {Calixto}}\ and\ \bibinfo {author} {\bibfnamefont {E.}~\bibnamefont
  {Romera}},\ }\bibfield  {title} {\bibinfo {title} {Identifying
  topological-band insulator transitions in silicene and other {2D} gapped
  dirac materials by means of {Rényi-Wehrl} entropy},\ }\href
  {https://doi.org/10.1209/0295-5075/109/40003} {\bibfield  {journal} {\bibinfo
   {journal} {{EPL} (Europhysics Letters)}\ }\textbf {\bibinfo {volume}
  {109}},\ \bibinfo {pages} {40003} (\bibinfo {year}
  {2015}{\natexlab{a}})}\BibitemShut {NoStop}%
\bibitem [{\citenamefont {Romera}\ and\ \citenamefont
  {Calixto}(2015)}]{silicene2}%
  \BibitemOpen
  \bibfield  {author} {\bibinfo {author} {\bibfnamefont {E.}~\bibnamefont
  {Romera}}\ and\ \bibinfo {author} {\bibfnamefont {M.}~\bibnamefont
  {Calixto}},\ }\bibfield  {title} {\bibinfo {title} {Uncertainty relations and
  topological-band insulator transitions in {2D} gapped dirac materials},\
  }\href {https://doi.org/10.1088/0953-8984/27/17/175003} {\bibfield  {journal}
  {\bibinfo  {journal} {Journal of Physics: Condensed Matter}\ }\textbf
  {\bibinfo {volume} {27}},\ \bibinfo {pages} {175003} (\bibinfo {year}
  {2015})}\BibitemShut {NoStop}%
\bibitem [{\citenamefont {Calixto}\ and\ \citenamefont
  {Romera}(2015{\natexlab{b}})}]{silicene3}%
  \BibitemOpen
  \bibfield  {author} {\bibinfo {author} {\bibfnamefont {M.}~\bibnamefont
  {Calixto}}\ and\ \bibinfo {author} {\bibfnamefont {E.}~\bibnamefont
  {Romera}},\ }\bibfield  {title} {\bibinfo {title} {Inverse participation
  ratio and localization in topological insulator phase transitions},\ }\href
  {https://doi.org/10.1088/1742-5468/2015/06/p06029} {\bibfield  {journal}
  {\bibinfo  {journal} {Journal of Statistical Mechanics: Theory and
  Experiment}\ }\textbf {\bibinfo {volume} {2015}},\ \bibinfo {pages} {P06029}
  (\bibinfo {year} {2015}{\natexlab{b}})}\BibitemShut {NoStop}%
\bibitem [{\citenamefont {Calixto}\ \emph {et~al.}(2021)\citenamefont
  {Calixto}, \citenamefont {Romera},\ and\ \citenamefont {Casta\~nos}}]{IJQC}%
  \BibitemOpen
  \bibfield  {author} {\bibinfo {author} {\bibfnamefont {M.}~\bibnamefont
  {Calixto}}, \bibinfo {author} {\bibfnamefont {E.}~\bibnamefont {Romera}},\
  and\ \bibinfo {author} {\bibfnamefont {O.}~\bibnamefont {Casta\~nos}},\
  }\bibfield  {title} {\bibinfo {title} {Analogies between the topological
  insulator phase of {2D} {Dirac} materials and the superradiant phase of
  atom-field systems},\ }\href {https://doi.org/10.1002/qua.26464} {\bibfield
  {journal} {\bibinfo  {journal} {International Journal of Quantum Chemistry}\
  }\textbf {\bibinfo {volume} {121}},\ \bibinfo {pages} {e26464} (\bibinfo
  {year} {2021})}\BibitemShut {NoStop}%
\bibitem [{\citenamefont {Castaños}\ \emph {et~al.}(2019)\citenamefont
  {Castaños}, \citenamefont {Romera},\ and\ \citenamefont {Calixto}}]{MRX}%
  \BibitemOpen
  \bibfield  {author} {\bibinfo {author} {\bibfnamefont {O.}~\bibnamefont
  {Castaños}}, \bibinfo {author} {\bibfnamefont {E.}~\bibnamefont {Romera}},\
  and\ \bibinfo {author} {\bibfnamefont {M.}~\bibnamefont {Calixto}},\
  }\bibfield  {title} {\bibinfo {title} {Information theoretic analysis of
  landau levels in monolayer phosphorene under magnetic and electric fields},\
  }\href {https://doi.org/10.1088/2053-1591/ab3fdc} {\bibfield  {journal}
  {\bibinfo  {journal} {Materials Research Express}\ }\textbf {\bibinfo
  {volume} {6}},\ \bibinfo {pages} {106316} (\bibinfo {year}
  {2019})}\BibitemShut {NoStop}%
\bibitem [{\citenamefont {Calixto}\ \emph {et~al.}(2022)\citenamefont
  {Calixto}, \citenamefont {Cordero}, \citenamefont {Romera},\ and\
  \citenamefont {Castaños}}]{CALIXTO2022128057}%
  \BibitemOpen
  \bibfield  {author} {\bibinfo {author} {\bibfnamefont {M.}~\bibnamefont
  {Calixto}}, \bibinfo {author} {\bibfnamefont {N.~A.}\ \bibnamefont
  {Cordero}}, \bibinfo {author} {\bibfnamefont {E.}~\bibnamefont {Romera}},\
  and\ \bibinfo {author} {\bibfnamefont {O.}~\bibnamefont {Castaños}},\
  }\bibfield  {title} {\bibinfo {title} {Signatures of topological phase
  transitions in higher {Landau} levels of {HgTe/CdTe} quantum wells from an
  information theory perspective},\ }\href
  {https://doi.org/10.1016/j.physa.2022.128057} {\bibfield  {journal} {\bibinfo
   {journal} {Physica A: Statistical Mechanics and its Applications}\ }\textbf
  {\bibinfo {volume} {605}},\ \bibinfo {pages} {128057} (\bibinfo {year}
  {2022})}\BibitemShut {NoStop}%
\bibitem [{\citenamefont {Calixto}\ \emph {et~al.}(2024)\citenamefont
  {Calixto}, \citenamefont {Mayorgas}, \citenamefont {Cordero}, \citenamefont
  {Romera},\ and\ \citenamefont {Castaños}}]{SciPostPhys.16.3.077}%
  \BibitemOpen
  \bibfield  {author} {\bibinfo {author} {\bibfnamefont {M.}~\bibnamefont
  {Calixto}}, \bibinfo {author} {\bibfnamefont {A.}~\bibnamefont {Mayorgas}},
  \bibinfo {author} {\bibfnamefont {N.~A.}\ \bibnamefont {Cordero}}, \bibinfo
  {author} {\bibfnamefont {E.}~\bibnamefont {Romera}},\ and\ \bibinfo {author}
  {\bibfnamefont {O.}~\bibnamefont {Castaños}},\ }\bibfield  {title} {\bibinfo
  {title} {Faraday rotation and transmittance as markers of topological phase
  transitions in 2d materials},\ }\href
  {https://doi.org/10.21468/SciPostPhys.16.3.077} {\bibfield  {journal}
  {\bibinfo  {journal} {SciPost Phys.}\ }\textbf {\bibinfo {volume} {16}},\
  \bibinfo {pages} {077} (\bibinfo {year} {2024})}\BibitemShut {NoStop}%
\bibitem [{\citenamefont {Calixto}\ and\ \citenamefont
  {Castaños}(2025)}]{IJMPB2025}%
  \BibitemOpen
  \bibfield  {author} {\bibinfo {author} {\bibfnamefont {M.}~\bibnamefont
  {Calixto}}\ and\ \bibinfo {author} {\bibfnamefont {O.}~\bibnamefont
  {Castaños}},\ }\bibfield  {title} {\bibinfo {title} {Localization and
  entanglement characterization of edge states in {HgTe} quantum wells in a
  finite strip geometry},\ }\href {https://doi.org/10.1142/S0217979225502637}
  {\bibfield  {journal} {\bibinfo  {journal} {International Journal of Modern
  Physics B}\ }\textbf {\bibinfo {volume} {39}},\ \bibinfo {pages} {2550263}
  (\bibinfo {year} {2025})}\BibitemShut {NoStop}%
\bibitem [{\citenamefont {Calixto}\ \emph {et~al.}(2025)\citenamefont
  {Calixto}, \citenamefont {Mayorgas},\ and\ \citenamefont
  {Castaños}}]{TBGPhysicaE}%
  \BibitemOpen
  \bibfield  {author} {\bibinfo {author} {\bibfnamefont {M.}~\bibnamefont
  {Calixto}}, \bibinfo {author} {\bibfnamefont {A.}~\bibnamefont {Mayorgas}},\
  and\ \bibinfo {author} {\bibfnamefont {O.}~\bibnamefont {Castaños}},\
  }\bibfield  {title} {\bibinfo {title} {Capturing magic angles in twisted
  bilayer graphene from information theory markers},\ }\href
  {https://doi.org/https://doi.org/10.1016/j.physe.2025.116199} {\bibfield
  {journal} {\bibinfo  {journal} {Physica E: Low-dimensional Systems and
  Nanostructures}\ }\textbf {\bibinfo {volume} {169}},\ \bibinfo {pages}
  {116199} (\bibinfo {year} {2025})}\BibitemShut {NoStop}%
\bibitem [{\citenamefont {Gu}(2010)}]{Gu}%
  \BibitemOpen
  \bibfield  {author} {\bibinfo {author} {\bibfnamefont {S.-J.}\ \bibnamefont
  {Gu}},\ }\bibfield  {title} {\bibinfo {title} {Fidelity approach to quantum
  phase transitions},\ }\href {https://doi.org/10.1142/S0217979210056335}
  {\bibfield  {journal} {\bibinfo  {journal} {International Journal of Modern
  Physics B}\ }\textbf {\bibinfo {volume} {24}},\ \bibinfo {pages}
  {4371–4458} (\bibinfo {year} {2010})}\BibitemShut {NoStop}%
\bibitem [{\citenamefont {Zanardi}\ and\ \citenamefont {{Paunkovi\ifmmode
  \acute{c}\else ć\fi{}}}(2006)}]{Zanardi}%
  \BibitemOpen
  \bibfield  {author} {\bibinfo {author} {\bibfnamefont {P.}~\bibnamefont
  {Zanardi}}\ and\ \bibinfo {author} {\bibfnamefont {N.}~\bibnamefont
  {{Paunkovi\ifmmode \acute{c}\else ć\fi{}}}},\ }\bibfield  {title} {\bibinfo
  {title} {Ground state overlap and quantum phase transitions},\ }\href
  {https://doi.org/10.1103/PhysRevE.74.031123} {\bibfield  {journal} {\bibinfo
  {journal} {Phys. Rev. E}\ }\textbf {\bibinfo {volume} {74}},\ \bibinfo
  {pages} {031123} (\bibinfo {year} {2006})}\BibitemShut {NoStop}%
\bibitem [{\citenamefont {Bistritzer}\ and\ \citenamefont
  {MacDonald}(2011)}]{Bistritzer12233}%
  \BibitemOpen
  \bibfield  {author} {\bibinfo {author} {\bibfnamefont {R.}~\bibnamefont
  {Bistritzer}}\ and\ \bibinfo {author} {\bibfnamefont {A.~H.}\ \bibnamefont
  {MacDonald}},\ }\bibfield  {title} {\bibinfo {title} {Moiré bands in twisted
  double-layer graphene},\ }\href {https://doi.org/10.1073/pnas.1108174108}
  {\bibfield  {journal} {\bibinfo  {journal} {Proceedings of the National
  Academy of Sciences}\ }\textbf {\bibinfo {volume} {108}},\ \bibinfo {pages}
  {12233–12237} (\bibinfo {year} {2011})}\BibitemShut {NoStop}%
\bibitem [{\citenamefont {Cao}\ \emph {et~al.}(2018{\natexlab{a}})\citenamefont
  {Cao}, \citenamefont {Fatemi}, \citenamefont {Demir}, \citenamefont {Fang},
  \citenamefont {Tomarken}, \citenamefont {Luo}, \citenamefont
  {Sanchez-Yamagishi}, \citenamefont {Watanabe}, \citenamefont {Taniguchi},
  \citenamefont {Kaxiras}, \citenamefont {Ashoori},\ and\ \citenamefont
  {Jarillo-Herrero}}]{Cao2018}%
  \BibitemOpen
  \bibfield  {author} {\bibinfo {author} {\bibfnamefont {Y.}~\bibnamefont
  {Cao}}, \bibinfo {author} {\bibfnamefont {V.}~\bibnamefont {Fatemi}},
  \bibinfo {author} {\bibfnamefont {A.}~\bibnamefont {Demir}}, \bibinfo
  {author} {\bibfnamefont {S.}~\bibnamefont {Fang}}, \bibinfo {author}
  {\bibfnamefont {S.~L.}\ \bibnamefont {Tomarken}}, \bibinfo {author}
  {\bibfnamefont {J.~Y.}\ \bibnamefont {Luo}}, \bibinfo {author} {\bibfnamefont
  {J.~D.}\ \bibnamefont {Sanchez-Yamagishi}}, \bibinfo {author} {\bibfnamefont
  {K.}~\bibnamefont {Watanabe}}, \bibinfo {author} {\bibfnamefont
  {T.}~\bibnamefont {Taniguchi}}, \bibinfo {author} {\bibfnamefont
  {E.}~\bibnamefont {Kaxiras}}, \bibinfo {author} {\bibfnamefont {R.~C.}\
  \bibnamefont {Ashoori}},\ and\ \bibinfo {author} {\bibfnamefont
  {P.}~\bibnamefont {Jarillo-Herrero}},\ }\bibfield  {title} {\bibinfo {title}
  {Correlated insulator behaviour at half-filling in magic-angle graphene
  superlattices},\ }\href {https://doi.org/10.1038/nature26154} {\bibfield
  {journal} {\bibinfo  {journal} {Nature}\ }\textbf {\bibinfo {volume} {556}},\
  \bibinfo {pages} {80–84} (\bibinfo {year}
  {2018}{\natexlab{a}})}\BibitemShut {NoStop}%
\bibitem [{\citenamefont {Cao}\ \emph {et~al.}(2018{\natexlab{b}})\citenamefont
  {Cao}, \citenamefont {Fatemi}, \citenamefont {Fang}, \citenamefont
  {Watanabe}, \citenamefont {Taniguchi}, \citenamefont {Kaxiras},\ and\
  \citenamefont {Jarillo-Herrero}}]{Cao2018unconventional}%
  \BibitemOpen
  \bibfield  {author} {\bibinfo {author} {\bibfnamefont {Y.}~\bibnamefont
  {Cao}}, \bibinfo {author} {\bibfnamefont {V.}~\bibnamefont {Fatemi}},
  \bibinfo {author} {\bibfnamefont {S.}~\bibnamefont {Fang}}, \bibinfo {author}
  {\bibfnamefont {K.}~\bibnamefont {Watanabe}}, \bibinfo {author}
  {\bibfnamefont {T.}~\bibnamefont {Taniguchi}}, \bibinfo {author}
  {\bibfnamefont {E.}~\bibnamefont {Kaxiras}},\ and\ \bibinfo {author}
  {\bibfnamefont {P.}~\bibnamefont {Jarillo-Herrero}},\ }\bibfield  {title}
  {\bibinfo {title} {Unconventional superconductivity in magic-angle graphene
  superlattices},\ }\href {https://doi.org/10.1038/nature26160} {\bibfield
  {journal} {\bibinfo  {journal} {Nature}\ }\textbf {\bibinfo {volume} {556}},\
  \bibinfo {pages} {43–50} (\bibinfo {year}
  {2018}{\natexlab{b}})}\BibitemShut {NoStop}%
\bibitem [{\citenamefont {Zhou}\ \emph {et~al.}(2008)\citenamefont {Zhou},
  \citenamefont {Lu}, \citenamefont {Chu}, \citenamefont {Shen},\ and\
  \citenamefont {Niu}}]{PhysRevLett.101.246807}%
  \BibitemOpen
  \bibfield  {author} {\bibinfo {author} {\bibfnamefont {B.}~\bibnamefont
  {Zhou}}, \bibinfo {author} {\bibfnamefont {H.-Z.}\ \bibnamefont {Lu}},
  \bibinfo {author} {\bibfnamefont {R.-L.}\ \bibnamefont {Chu}}, \bibinfo
  {author} {\bibfnamefont {S.-Q.}\ \bibnamefont {Shen}},\ and\ \bibinfo
  {author} {\bibfnamefont {Q.}~\bibnamefont {Niu}},\ }\bibfield  {title}
  {\bibinfo {title} {Finite size effects on helical edge states in a quantum
  spin-hall system},\ }\href {https://doi.org/10.1103/PhysRevLett.101.246807}
  {\bibfield  {journal} {\bibinfo  {journal} {Phys. Rev. Lett.}\ }\textbf
  {\bibinfo {volume} {101}},\ \bibinfo {pages} {246807} (\bibinfo {year}
  {2008})}\BibitemShut {NoStop}%
\bibitem [{\citenamefont {Fang}\ \emph {et~al.}(2025)\citenamefont {Fang},
  \citenamefont {Yang}, \citenamefont {Wang}, \citenamefont {Yang},
  \citenamefont {Lu}, \citenamefont {Lee}, \citenamefont {Wang},\ and\
  \citenamefont {Ang}}]{fang2025edgetronicstwodimensionalaltermagnets}%
  \BibitemOpen
  \bibfield  {author} {\bibinfo {author} {\bibfnamefont {S.}~\bibnamefont
  {Fang}}, \bibinfo {author} {\bibfnamefont {Z.}~\bibnamefont {Yang}}, \bibinfo
  {author} {\bibfnamefont {J.}~\bibnamefont {Wang}}, \bibinfo {author}
  {\bibfnamefont {X.}~\bibnamefont {Yang}}, \bibinfo {author} {\bibfnamefont
  {J.}~\bibnamefont {Lu}}, \bibinfo {author} {\bibfnamefont {C.~H.}\
  \bibnamefont {Lee}}, \bibinfo {author} {\bibfnamefont {X.}~\bibnamefont
  {Wang}},\ and\ \bibinfo {author} {\bibfnamefont {Y.~S.}\ \bibnamefont
  {Ang}},\ }\href {https://arxiv.org/abs/2508.10451} {\bibinfo {title}
  {Edgetronics in two-dimensional altermagnets}} (\bibinfo {year} {2025}),\
  \Eprint {https://arxiv.org/abs/2508.10451} {arXiv:2508.10451
  [cond-mat.mes-hall]} \BibitemShut {NoStop}%
\bibitem [{\citenamefont {Chen}\ \emph {et~al.}(2025)\citenamefont {Chen},
  \citenamefont {Liu}, \citenamefont {Lu},\ and\ \citenamefont
  {Xie}}]{ChenPRL135_2025}%
  \BibitemOpen
  \bibfield  {author} {\bibinfo {author} {\bibfnamefont {Y.}~\bibnamefont
  {Chen}}, \bibinfo {author} {\bibfnamefont {X.}~\bibnamefont {Liu}}, \bibinfo
  {author} {\bibfnamefont {H.-Z.}\ \bibnamefont {Lu}},\ and\ \bibinfo {author}
  {\bibfnamefont {X.~C.}\ \bibnamefont {Xie}},\ }\bibfield  {title} {\bibinfo
  {title} {Electrical switching of altermagnetism},\ }\href
  {https://doi.org/10.1103/zm5y-vy41} {\bibfield  {journal} {\bibinfo
  {journal} {Phys. Rev. Lett.}\ }\textbf {\bibinfo {volume} {135}},\ \bibinfo
  {pages} {016701} (\bibinfo {year} {2025})}\BibitemShut {NoStop}%
\bibitem [{\citenamefont {Parshukov}\ \emph {et~al.}(2025)\citenamefont
  {Parshukov}, \citenamefont {Wiedmann},\ and\ \citenamefont
  {Schnyder}}]{Parshukov_PhysRevB.111.224406}%
  \BibitemOpen
  \bibfield  {author} {\bibinfo {author} {\bibfnamefont {K.}~\bibnamefont
  {Parshukov}}, \bibinfo {author} {\bibfnamefont {R.}~\bibnamefont
  {Wiedmann}},\ and\ \bibinfo {author} {\bibfnamefont {A.~P.}\ \bibnamefont
  {Schnyder}},\ }\bibfield  {title} {\bibinfo {title} {Topological crossings in
  two-dimensional altermagnets: Symmetry classification and topological
  responses},\ }\href {https://doi.org/10.1103/PhysRevB.111.224406} {\bibfield
  {journal} {\bibinfo  {journal} {Phys. Rev. B}\ }\textbf {\bibinfo {volume}
  {111}},\ \bibinfo {pages} {224406} (\bibinfo {year} {2025})}\BibitemShut
  {NoStop}%
\bibitem [{\citenamefont {Jungwirth}\ \emph {et~al.}(2025)\citenamefont
  {Jungwirth}, \citenamefont {Fernandes}, \citenamefont {Fradkin},
  \citenamefont {MacDonald}, \citenamefont {Sinova},\ and\ \citenamefont
  {Šmejkal}}]{JUNGWIRTH2025100162}%
  \BibitemOpen
  \bibfield  {author} {\bibinfo {author} {\bibfnamefont {T.}~\bibnamefont
  {Jungwirth}}, \bibinfo {author} {\bibfnamefont {R.~M.}\ \bibnamefont
  {Fernandes}}, \bibinfo {author} {\bibfnamefont {E.}~\bibnamefont {Fradkin}},
  \bibinfo {author} {\bibfnamefont {A.~H.}\ \bibnamefont {MacDonald}}, \bibinfo
  {author} {\bibfnamefont {J.}~\bibnamefont {Sinova}},\ and\ \bibinfo {author}
  {\bibfnamefont {L.}~\bibnamefont {Šmejkal}},\ }\bibfield  {title} {\bibinfo
  {title} {Altermagnetism: An unconventional spin-ordered phase of matter},\
  }\href {https://doi.org/https://doi.org/10.1016/j.newton.2025.100162}
  {\bibfield  {journal} {\bibinfo  {journal} {Newton}\ }\textbf {\bibinfo
  {volume} {1}},\ \bibinfo {pages} {100162} (\bibinfo {year}
  {2025})}\BibitemShut {NoStop}%
\bibitem [{\citenamefont {Sicheler}\ \emph {et~al.}(2025)\citenamefont
  {Sicheler}, \citenamefont {Raimondi}, \citenamefont {Sangiovanni},\ and\
  \citenamefont {Re}}]{sicheler2025opticallytunablespintransport}%
  \BibitemOpen
  \bibfield  {author} {\bibinfo {author} {\bibfnamefont {N.}~\bibnamefont
  {Sicheler}}, \bibinfo {author} {\bibfnamefont {R.}~\bibnamefont {Raimondi}},
  \bibinfo {author} {\bibfnamefont {G.}~\bibnamefont {Sangiovanni}},\ and\
  \bibinfo {author} {\bibfnamefont {L.~D.}\ \bibnamefont {Re}},\ }\href
  {https://arxiv.org/abs/2508.06938} {\bibinfo {title} {Optically tunable spin
  transport in bilayer altermagnetic mott insulators}} (\bibinfo {year}
  {2025}),\ \Eprint {https://arxiv.org/abs/2508.06938} {arXiv:2508.06938
  [cond-mat.str-el]} \BibitemShut {NoStop}%
\end{thebibliography}%

\end{document}